# An Overview of Lead and Accompaniment Separation in Music

Zafar Rafii, *Member, IEEE,* Antoine Liutkus, *Member, IEEE,* Fabian-Robert Stöter,
Stylianos Ioannis Mimilakis, *Student Member, IEEE,* Derry FitzGerald, and Bryan Pardo, *Member, IEEE*

*Abstract*—Popular music is often composed of an accompaniment and a *lead* component, the latter typically consisting of vocals. Filtering such mixtures to extract one or both components has many applications, such as automatic karaoke and remixing. This particular case of *source separation* yields very specific challenges and opportunities, including the particular complexity of musical structures, but also relevant prior knowledge coming from acoustics, musicology or sound engineering. Due to both its importance in applications and its challenging difficulty, lead and accompaniment separation has been a popular topic in signal processing for decades. In this article, we provide a comprehensive review of this research topic, organizing the different approaches according to whether they are model-based or data-centered. For model-based methods, we organize them according to whether they concentrate on the lead signal, the accompaniment, or both. For data-centered approaches, we discuss the particular difficulty of obtaining data for learning lead separation systems, and then review recent approaches, notably those based on deep learning. Finally, we discuss the delicate problem of evaluating the quality of music separation through adequate metrics and present the results of the largest evaluation, to-date, of lead and accompaniment separation systems. In conjunction with the above, a comprehensive list of references is provided, along with relevant pointers to available implementations and repositories.

## I. INTRODUCTION

**M**USIC is a major form of artistic expression and plays a central role in the entertainment industry. While digitization and the Internet led to a revolution in the way music reaches its audience [1], [2], there is still much room to improve on how one interacts with musical content, beyond simply controlling the master volume and equalization. The ability to interact with the individual audio objects (e.g., the lead vocals) in a music recording would enable diverse applications such as music upmixing and remixing, automatic karaoke, object-wise equalization, etc.

Most publicly available music recordings (e.g., CDs, YouTube, iTunes, Spotify) are distributed as mono or stereo

Z. Rafii is with Gracenote, Emeryville, CA, USA (zafar.rafii@nielsen.com). A. Liutkus and F.-R. Stöter are with Inria and LIRMM, University of Montpellier, France (firstname.lastname@inria.fr). S.I. Mimilakis is with Fraunhofer IDMT, Ilmenau, Germany (mis@idmt.fraunhofer.de). D. FitzGerald is with Cork School of Music, Cork Institute of Technology, Cork, Ireland (Derry.Fitzgerald@cit.ie). B. Pardo is with Northwestern University, Evanston, IL, USA (pardo@northwestern.edu).

This work was partly supported by the research programme KAMoulox (ANR-15-CE38-0003-01) funded by ANR, the French State agency for research. S.I. Mimilakis is supported by the European Unions H2020 Framework Programme (H2020- MSCA-ITN-2014) under grant agreement no 642685 MacSeNet mixtures with multiple sound objects sharing a track. Therefore, manipulation of individual sound objects requires separation of the stereo audio mixture into several tracks, one for each different sound sources. This process is called *audio source separation* and this overview paper is concerned with an important particular case: isolating the *lead* source — typically, the vocals — from the musical accompaniment (all the rest of the signal).

As a general problem in applied mathematics, source separation has enjoyed tremendous research activity for roughly 50 years and has applications in various fields such as bioinformatics, telecommunications, and audio. Early research focused on so-called *blind* source separation, which typically builds on very weak assumptions about the signals that comprise the mixture in conjunction with very strong assumptions on the way they are mixed. The reader is referred to [3], [4] for a comprehensive review on blind source separation. Typical blind algorithms, e.g., independent component analysis (ICA) [5], [6], depend on assumptions such as: source signals are independent, there are more mixture channels than there are signals, and mixtures are well modeled as a linear combination of signals. While such assumptions are appropriate for some signals like electroencephalograms, they are often violated in audio.

Much research in audio-specific source separation [7], [8] has been motivated by the *speech enhancement* problem [9], which aims to recover clean speech from noisy recordings and can be seen as a particular instance of source separation. In this respect, many algorithms assume the audio background can be modeled as stationary. However, the musical sources are characterized by a very rich, non-stationary spectro-temporal structure. This prohibits the use of such methods. Musical sounds often exhibit highly synchronous evolution over both time and frequency, making overlap in both time and frequency very common. Furthermore, a typical commercial music mixture violates all the classical assumptions of ICA. Instruments are correlated (e.g., a chorus of singers), there are more instruments than channels in the mixture, and there are non-linearities in the mixing process (e.g., dynamic range compression). This all has required the development of music-specific algorithms, exploiting available prior information about source structure or mixing parameters [10], [11].

This article provides an overview of nearly 50 years of research on lead and accompaniment separation in music. Due to space constraints and the large variability of the paradigms involved, we cannot delve into detailed mathematical description of each method. Instead, we will convey core ideas and methodologies, grouping approaches according to common features. As with any attempt to impose an *a posteriori* taxonomy on such a large body of research, the resulting classification is arguable. However, we believe it is useful as a roadmap of the relevant literature.

Our objective is not to advocate one methodology over another. While the most recent methods — in particular those based on deep learning — currently show the best performance, we believe that ideas underlying earlier methods may also be inspiring and stimulate new research. This point of view leads us to focus more on the strengths of the methods rather than on their weaknesses.

The rest of the article is organized as follows. In Section II, we present the basic concepts needed to understand the discussion. We then present sections on *model-based* methods that exploit specific knowledge about the lead and/or the accompaniment signals in music to achieve separation. We show in Section III how one body of research is focused on modeling the lead signal as harmonic, exploiting this central assumption for separation. Then, Section IV describes many methods achieving separation using a model that takes the musical accompaniment as *redundant*. In Section V, we show how these two ideas were combined in other studies to achieve separation. Then, we present data-driven approaches in Section VI, which exploit large databases of audio examples where both the isolated lead and accompaniment signals are available. This enables the use of machine learning methods to learn how to separate. In Section VII, we show how the widespread availability of stereo signals may be leveraged to design algorithms that assume centered-panned vocals, but also to improve separation of most methods. Finally, Section VIII is concerned with the problem of how to evaluate the quality of the separation, and provides the results for the largest evaluation campaign to date on this topic.

## II. FUNDAMENTAL CONCEPTS

We now very briefly describe the basic ideas required to understand this paper, classified into three main categories: signal processing, audio modeling and probability theory. The interested reader is strongly encouraged to delve into the many online courses or textbooks available for a more detailed presentation of these topics, such as [12], [13] for signal processing, [9] for speech modeling, and [14], [15] for probability theory.

### A. Signal processing

Sound is a series of pressure waves in the air. It is recorded as a *waveform*, a time-series of measurements of the displacement of the microphone diaphragm in response to these pressure waves. Sound is reproduced if a loudspeaker diaphragm is moved according to the recorded waveform. Multichannel signals simply consist of several waveforms, captured by more than one microphone. Typically, music signals are stereophonic, containing two waveforms.

Microphone displacement is typically measured at a fixed *sampling frequency*. In music processing, it is common to have sampling frequencies of $44.1$ kHz (the sample frequency on a compact disc) or $48$ kHz, which are higher than the typical sampling rates of 16 kHz or 8 kHz used for speech in telephony. This is because musical signals contain much higher frequency content than speech and the goal is aesthetic beauty in addition to basic intelligibility.

A time-frequency (TF) representation of sound is a matrix that encodes the time-varying *spectrum* of the waveform. Its entries are called TF *bins* and encode the varying spectrum of the waveform for all time frames and frequency channels. The most commonly-used TF representation is the short time Fourier transform (STFT) [16], which has complex entries: the angle accounts for the phase, i.e., the actual shift of the corresponding sinusoid at that time bin and frequency bin, and the magnitude accounts for the amplitude of that sinusoid in the signal. The magnitude (or power) of the STFT is called *spectrogram*. When the mixture is multichannel, the TF representation for each channel is computed, leading to a three-dimensional array: frequency, time and channel.

A TF representation is typically used as a first step in processing the audio because sources tend to be less overlapped in the TF representation than in the waveform [17]. This makes it easier to select portions of a mixture that correspond to only a single source. An STFT is typically used because it can be inverted back to the original waveform. Therefore, modifications made to the STFT can be used to create a modified waveform. Generally, a linear mixing process is considered, i.e., the mixture signal is equal to the sum of the source signals. Since the Fourier transform is a linear operation, this equality holds for the STFT. While that is not the case for the magnitude (or power) of the STFT, it is commonly assumed that the spectrograms of the sources sum to the spectrogram of the mixture.

In many methods, the separated sources are obtained by *filtering* the mixture. This can be understood as performing some equalization on the mixture, where each frequency is attenuated or kept intact. Since both the lead and the accompaniment signals change over time, the filter also changes. This is typically done using a TF *mask*, which, in its simplest form, is defined as the gain between $0$ and $1$ to apply on each element of the TF representation of the mixture (e.g., an STFT) in order to estimate the desired signal. Loosely speaking, it can be understood as an equalizer whose setting changes every few milliseconds. After multiplication of the mixture by a mask, the separated signal is recovered through an inverse TF transform. In the multichannel setting, more sophisticated filters may be designed that incorporate some delay and combine different channels; this is usually called *beamforming*. In the frequency domain, this is often equivalent to using complex matrices to multiply the mixture TF representation with, instead of just scalars between $0$ and $1$.

In practice, masks can be designed to filter the mixture in several ways. One may estimate the spectrogram for a single source or component, e.g., the accompaniment, and subtract it from the mixture spectrogram, e.g., in order to estimate the lead [18]. Another way would be to estimate separate spectrograms for both lead and accompaniment and combine them to yield a mask. For instance, a TF mask for the lead can be taken as the proportion of the lead spectrogram over the sum of both spectrograms, at each TF bin. Such filters are often called *Wiener filters* [19] or *ratio masks*. How they are calculated may involve some additional techniques like exponentiation and may be understood according to assumptions regarding the underlying statistics of the sources. For recent work in this area, and many useful pointers in designing such masks, the reader is referred to [20].

### B. Audio and speech modeling

It is typical in audio processing to describe audio waveforms as belonging to one of two different categories, which are *sinusoidal signals* — or pure tones — and *noise*. Actually, both are just the two extremes in a continuum of varying *predictability*: on the one hand, the shape of a sinusoidal wave in the future can reliably be guessed from previous samples. On the other hand, white noise is *defined* as an unpredictable signal and its spectrogram has constant energy everywhere. Different noise profiles may then be obtained by attenuating the energy of some frequency regions. This in turn induces some predictability in the signal, and in the extreme case where all the energy content is concentrated in one frequency, a pure tone is obtained.

A waveform may always be modeled as some *filter* applied on some *excitation signal*. Usually, the filter is assumed to vary smoothly across frequencies, hence modifying only what is called *the spectral envelope* of the signal, while the excitation signal comprises the rest. This is the basis for the *source-filter* model [21], which is of great importance in speech modeling, and thus also in vocal separation. As for speech, the filter is created by the shape of the vocal tract. The excitation signal is made of the glottal pulses generated by the vibration of the vocal folds. This results into *voiced* speech sounds made of time-varying harmonic/sinusoidal components. The excitation signal can also be the air flow passing through some constriction of the vocal tract. This results into *unvoiced*, noise-like, speech sounds. In this context, vowels are said to be voiced and tend to feature many sinusoids, while some phonemes such as fricatives are unvoiced and noisier.

A classical tool for dissociating the envelope from the excitation is the *cepstrum* [22]. It has applications for estimating the fundamental frequency [23], [24], for deriving the Mel-frequency cepstral coefficients (MFCC) [25], or for filtering signals through a so-called *liftering* operation [26] that enables modifications of either the excitation or the envelope parts through the source-filter paradigm.

An advantage of the source-filter model approach is indeed that one can dissociate the pitched content of the signal, embodied by the position of its harmonics, from its TF envelope which describes where the energy of the sound lies. In the case of vocals, it yields the ability to distinguish between the actual note being sung (pitch content) and the phoneme being uttered (mouth and vocal tract configuration), respectively. One key feature of vocals is they typically exhibit great variability in fundamental frequency over time. They can also exhibit larger *vibratos* (fundamental frequency modulations) and *tremolos* (amplitude modulations) in comparison to other instruments, as seen in the top spectrogram in Figure 1.

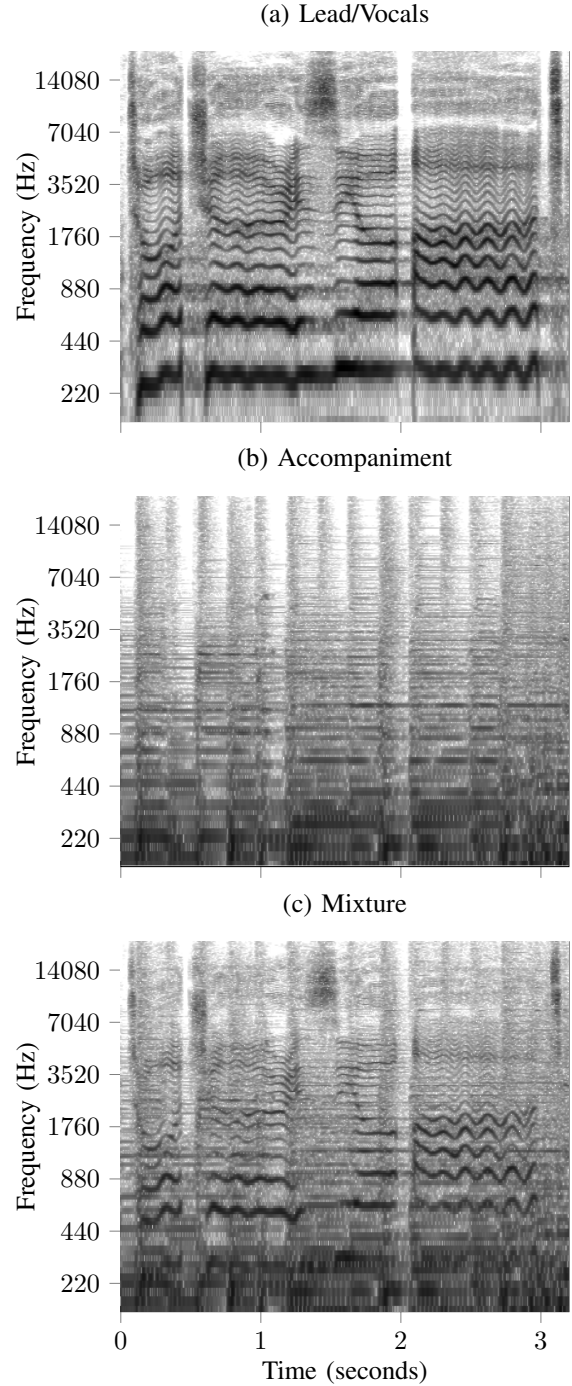

Fig. 1: Examples of spectrograms from an excerpt of the track "The Wrong'Uns - Rothko" from MUSDB18 dataset. The two sources to be separated are depicted in (a) and (b), and its mixture in (c). The vocals (a) are mostly harmonic and often well described by a source-filter model in which an excitation signal is filtered according to the vocal tract configuration. The accompaniment signal (b) features more diversity, but usually does not feature as much vibrato as for the vocals, and most importantly is seen to be *denser* and also *more redundant*. All spectrograms have log-compressed amplitudes as well as log-scaled frequency axis.

A particularity of musical signals is that they typically consist of sequences of pitched notes. A sound gives the perception of having a pitch if the majority of the energy in the audio signal is at frequencies located at integer multiples of some fundamental frequency. These integer multiples are called *harmonics*. When the fundamental frequency changes, the frequencies of these harmonics also change, yielding the typical comb spectrograms of harmonic signals, as depicted in the top spectrogram in Figure 1. Another noteworthy feature of sung melodies over simple speech is that their fundamental frequencies are, in general, located at precise frequency values corresponding to the musical key of the song. These very peculiar features are often exploited in separation methods. For simplicity reasons, we use the terms *pitch* and *fundamental frequency* interchangeably throughout the paper.

### C. Probability theory

Probability theory [14], [27] is an important framework for designing many data analysis and processing methods. Many of the methods described in this article use it and it is far beyond the scope of this paper to present it rigorously. For our purpose, it will suffice to say that the *observations* consist of the mixture signals. On the other hand, the *parameters* are any relevant feature about the source signal (such as pitch or time-varying envelope) or how the signals are mixed (e.g., the panning position). These parameters can be used to derive estimates about the target lead and accompaniment signals.

We understand a probabilistic *model* as a function of both the observations and the parameters: it describes how likely the observations are, given the parameters. For instance, a flat spectrum is likely under the noise model, and a mixture of comb spectrograms is likely under a harmonic model with the appropriate pitch parameters for the sources. When the observations are given, variation in the model depends only on the parameters. For some parameter value, it tells how likely the observations are. Under a harmonic model for instance, pitch may be estimated by finding the pitch parameter that makes the observed waveform as likely as possible. Alternatively, we may want to choose between several possible models such as voiced or unvoiced. In such cases, *model selection* methods are available, such as the Bayesian information criterion (BIC) [28].

Given these basic ideas, we briefly mention two models that are of particular importance. Firstly, the hidden Markov model (HMM) [15], [29] is relevant for time-varying observations. It basically defines several *states*, each one related to a specific model and with some probabilities for transitions between them. For instance, we could define as many states as possible notes played by the lead guitar, each one associated with a typical spectrum. The *Viterbi algorithm* is a dynamic programming method which actually estimates the most likely sequence of states given a sequence of observations [30]. Secondly, the Gaussian mixture model (GMM) [31] is a way to approximate any distribution as a weighted sum of Gaussians. It is widely used in clustering, because it works well with the celebrated Expectation-Maximization (EM) algorithm [32] to assign one particular cluster to each data point, while automatically estimating the clusters parameters. As we will see later, many methods work by assigning each TF bin to a given source in a similar way.

## III. Modeling the lead signal: harmonicity

As mentioned in Section II-B, one particularity of vocals is their production by the vibration of the vocal folds, further filtered by the vocal tract. As a consequence, sung melodies are *mostly* harmonic, as depicted in Figure 1, and therefore have a fundamental frequency. If one can track the pitch of the vocals, one can then estimate the energy at the harmonics of the fundamental frequency and reconstruct the voice. This is the basis of the oldest methods (as well as some more recent methods) we are aware of for separating the lead signal from a musical mixture.

Such methods are summarized in Figure 2. In a first step, the objective is to get estimates of the time-varying fundamental frequency for the lead at each time frame. A second step in this respect is then to track this fundamental frequency over time, in other words, to find the best sequence of estimates, in order to identify the melody line. This can done either by a suitable pitch detection method, or by exploiting the availability of the score. Such algorithms typically assume that the lead corresponds to the harmonic signal with strongest amplitude. For a review on the particular topic of melody extraction, the reader is referred to [33].

From this starting point, we can distinguish between two kinds of approaches, depending on how they exploit the pitch information.

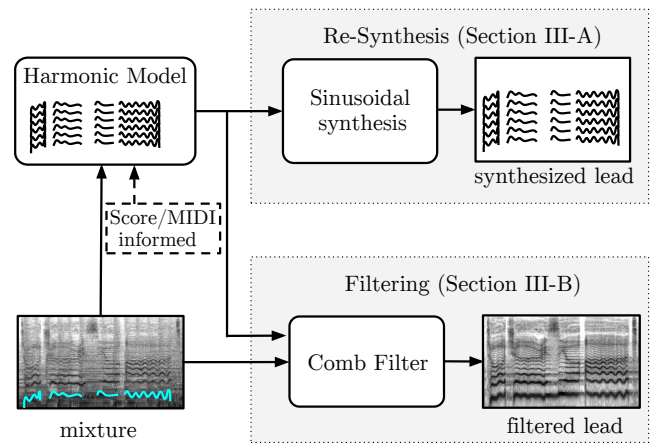

Fig. 2: The approaches based on a *harmonic assumption* for vocals. In a first analysis step, the fundamental frequency of the lead signal is extracted. From it, a separation is obtained either by resynthesis (Section III-A), or by filtering the mixture (Section III-B).

### A. Analysis-synthesis approaches

The first option to obtain the separated lead signal is to resynthesize it using a sinusoidal model. A sinusoidal model decomposes the sound with a set of sine waves of varying frequency and amplitude. If one knows the fundamental frequency of a pitched sound (like a singing voice), as well as the spectral envelope of the recording, then one can reconstruct the sound by making a set of sine waves whose frequencies are those of the harmonics of the fundamental frequency, and whose amplitudes are estimated from the spectral envelope of the audio. While the spectral envelope of the recording is generally not exactly the same as the spectral envelope of the target source, it can be a reasonable approximation, especially assuming that different sources do not overlap too much with each other in the TF representation of the mixture.

This idea allows for time-domain processing and was used in the earliest methods we are aware of. In 1973, Miller proposed in [34] to use the homomorphic vocoder [35] to separate the excitation function and impulse response of the vocal tract. Further refinements include segmenting parts of the signal as voiced, unvoiced, or silences using a heuristic program and manual interaction. Finally, cepstral liftering [26] was exploited to compensate for the noise or accompaniment.

Similarly, Maher used an analysis-synthesis approach in [36], assuming the mixtures are composed of only two harmonic sources. In his case, pitch detection was performed on the STFT and included heuristics to account for possibly colliding harmonics. He finally resynthesized each musical voice with a sinusoidal model.

Wang proposed instantaneous and frequency-warped techniques for signal parameterization and source separation, with application to voice separation in music [37], [38]. He introduced a frequency-locked loop algorithm which uses multiple harmonically constrained trackers. He computed the estimated fundamental frequency from a maximum-likelihood weighting of the tracking estimates. He was then able to estimate harmonic signals such as voices from complex mixtures.

Meron and Hirose proposed to separate singing voice and piano accompaniment [39]. In their case, prior knowledge consisting of musical scores was considered. Sinusoidal modeling as described in [40] was used.

Ben-Shalom and Dubnov proposed to filter an instrument or a singing voice out in such a way [41]. They first used a score alignment algorithm [42], assuming a known score. Then, they used the estimated pitch information to design a filter based on a harmonic model [43] and performed the filtering using the linear constraint minimum variance approach [44]. They additionally used a heuristic to deal with the unvoiced parts of the singing voice.

Zhang and Zhang proposed an approach based on harmonic structure modeling [45], [46]. They first extracted harmonic structures for singing voice and background music signals using a sinusoidal model [43], by extending the pitch estimation algorithm in [47]. Then, they used the clustering algorithm in [48] to learn harmonic structure models for the background music signals. Finally, they extracted the harmonic structures for all the instruments to reconstruct the background music signals and subtract them from the mixture, leaving only the singing voice signal.

More recently, Fujihara et al. proposed an accompaniment reduction method for singer identification [51]. After fundamental frequency estimation using [51], they extracted the harmonic structure of the melody, i.e., the power and phase of the sinusoidal components at fundamental frequency and harmonics. Finally, they resynthesized the audio signal of the melody using the sinusoidal model in [52].

Similarly, Mesaros et al. proposed a vocal separation method to help with singer identification [53]. They first applied a melody transcription system [54] which estimates the melody line with the corresponding MIDI note numbers. Then, they performed sinusoidal resynthesis, estimating amplitudes and phases from the polyphonic signal.

In a similar manner, Duan et al. proposed to separate harmonic sources, including singing voices, by using harmonic structure models [55]. They first defined an average harmonic structure model for an instrument. Then, they learned a model for each source by detecting the spectral peaks using a cross-correlation method [56] and quadratic interpolation [57]. Then, they extracted the harmonic structures using BIC and a clustering algorithm [48]. Finally, they separated the sources by re-estimating the fundamental frequencies, re-extracting the harmonics, and reconstructing the signals using a phase generation method [58].

Lagrange et al. proposed to formulate lead separation as a graph partition problem [59], [60]. They first identified peaks in the spectrogram and grouped the peaks into clusters by using a similarity measure which accounts for harmonically related peaks, and the normalized cut criterion [61] which is used for segmenting graphs in computer vision. They finally selected the cluster of peaks which corresponds to a predominant harmonic source and resynthesized it using a bank of sinusoidal oscillators.

Ryynänen et al. proposed to separate accompaniment from polyphonic music using melody transcription for karaoke applications [62]. They first transcribed the melody into a MIDI note sequence and a fundamental frequency trajectory, using the method in [63], an improved version of the earlier method [54]. Then, they used sinusoidal modeling to estimate, resynthesize, and remove the lead vocals from the musical mixture, using the quadratic polynomial-phase model in [64].

### B. Comb-filtering approaches

Using sinusoidal synthesis to generate the lead signal suffers from a typical *metallic* sound quality, which is mostly due to discrepancies between the estimated excitation signals of the lead signal compared to the ground truth. To address this issue, an alternative approach is to exploit harmonicity in another way, by filtering out everything from the mixture that is not located close to the detected harmonics.

Li and Wang proposed to use a vocal/non-vocal classifier and a predominant pitch detection algorithm [65], [66]. They first detected the singing voice by using a spectral change detector [67] to partition the mixture into homogeneous portions, and GMMs on MFCCs to classify the portions as vocal or non-vocal. Then, they used the predominant pitch detection algorithm in [68] to detect the pitch contours from the vocal portions, extending the multi-pitch tracking algorithm in [69]. Finally, they extracted the singing voice by decomposing the vocal portions into TF units and labeling them as singing or accompaniment dominant, extending the speech separation algorithm in [70].

Han and Raphael proposed an approach for desoloing a recording of a soloist with an accompaniment given a musical score and its time alignment with the recording [71]. They derived a mask [72] to remove the solo part after using an EM algorithm to estimate its melody, that exploits the score as side information.

Hsu et al. proposed an approach which also identifies and separates the unvoiced singing voice [73], [74]. Instead of processing in the STFT domain, they use the perceptually motivated gammatone filter-bank as in [66], [70]. They first detected accompaniment, unvoiced, and voiced segments using an HMM and identified voice-dominant TF units in the voiced frames by using the singing voice separation method in [66], using the predominant pitch detection algorithm in [75]. Unvoiced-dominant TF units were identified using a GMM classifier with MFCC features learned from training data. Finally, filtering was achieved with spectral subtraction [76].

Raphael and Han then proposed a classifier-based approach to separate a soloist from accompanying instruments using a time-aligned symbolic musical score [77]. They built a tree-structured classifier [78] learned from labeled training data to classify TF points in the STFT as belonging to solo or accompaniment. They additionally constrained their classifier to estimate masks having a connected structure.

Cano et al. proposed various approaches for solo and accompaniment separation. In [79], they separated saxophone melodies from mixtures with piano and/or orchestra by using a melody line detection algorithm, incorporating information about typical saxophone melody lines. In [80]–[82], they proposed to use the pitch detection algorithm in [83]. Then, they refined the fundamental frequency and the harmonics, and created a binary mask for the solo and accompaniment. They finally used a post-processing stage to refine the separation. In [84], they included a noise spectrum in the harmonic refinement stage to also capture noise-like sounds in vocals. In [85], they additionally included common amplitude modulation characteristics in the separation scheme.

Bosch et al. proposed to separate the lead instrument using a musical score [86]. After a preliminary alignment of the score to the mixture, they estimated a score confidence measure to deal with local misalignments and used it to guide the predominant pitch tracking. Finally, they performed low-latency separation based on the method in [87], by combining harmonic masks derived from the estimated pitch and additionally exploiting stereo information as presented later in Section VII.

Vaneph et al. proposed a framework for vocal isolation to help spectral editing [88]. They first used a voice activity detection process based on a deep learning technique [89]. Then, they used pitch tracking to detect the melodic line of the vocal and used it to separate the vocal and background, allowing a user to provide manual annotations when necessary.

## C. Shortcomings

As can be seen, explicitly assuming that the lead signal is harmonic led to an important body of research. While the aforementioned methods show excellent performance when their assumptions are valid, their performance can drop significantly in adverse, but common situations.

Firstly, vocals are not always purely harmonic as they contain unvoiced phonemes that are not harmonic. As seen above, some methods already handle this situation. However, vocals can also be whispered or saturated, both of which are difficult to handle with a harmonic model.

Secondly, methods based on the harmonic model depend on the quality of the pitch detection method. If the pitch detector switches from following the pitch of the lead (e.g., the voice) to another instrument, the wrong sound will be isolated from the mix. Often, pitch detectors assume the lead signal is the *loudest* harmonic sound in the mix. Unfortunately, this is not always the case. Another instrument may be louder or the lead may be silent for a passage. The tendency to follow the pitch of the wrong instrument can be mitigated by applying constraints on the pitch range to estimate and by using a perceptually relevant weighting filter before performing pitch tracking. Of course, these approaches do not help when the lead signal is silent.

## IV. Modeling the accompaniment: redundancy

In the previous section, we presented methods whose main focus was the modeling of a harmonic lead melody. Most of these studies did not make modeling the accompaniment a core focus. On the contrary, it was often dealt with as adverse noise to which the harmonic processing method should be robust to.

In this section, we present another line of research which concentrates on modeling the accompaniment under the assumption it is somehow more *redundant* than the lead signal. This assumption stems from the fact that musical accompaniments are often highly structured, with elements being repeated many times. Such repetitions can occur at the note level, in terms of rhythmic structure, or even from a harmonic point of view: instrumental notes are often constrained to have their pitch lie in a small set of frequencies. Therefore, modeling and removing the redundant elements of the signal are assumed to result in removal of the accompaniment.

In this paper, we identify three families of methods that exploit the redundancy of the accompaniment for separation.

### A. Grouping low-rank components

The first set of approaches we consider is the identification of redundancy in the accompaniment through the assumption that its spectrogram may be well represented by only a few components. Techniques exploiting this idea then focus on algebraic methods that decompose the mixture spectrogram into the product of a few template spectra activated over time. One way to do so is via non-negative matrix factorization (NMF) [90], [91], which incorporates non-negative constraints. In Figure 3, we picture methods exploiting such techniques. After factorization, we obtain several spectra, along with their activations over time. A subsequent step is the clustering of these spectra (and activations) into the lead or the accompaniment. Separation is finally performed by deriving Wiener filters to estimate the lead and the accompaniment from the mixture. For related applications of NMF in music analysis, the reader is referred to [92]–[94].

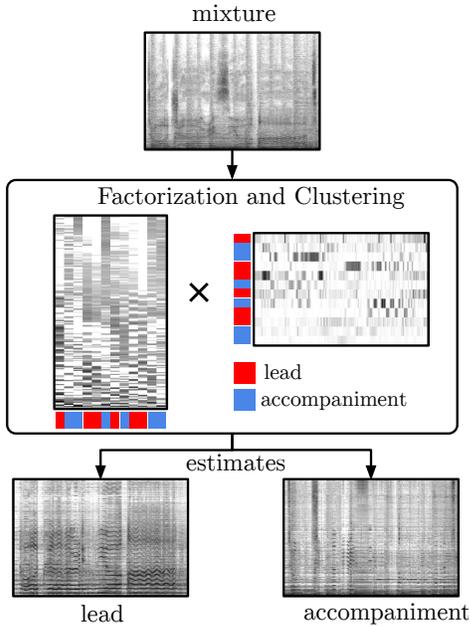

**Fig. 3:** The approaches based on a *low-rank* assumption. Non-negative matrix factorization (NMF) is used to identify *components* from the mixture, that are subsequently clustered into lead or accompaniment. Additional constraints may be incorporated.

Vembu and Baumann proposed to use NMF (and also ICA [95]) to separate vocals from mixtures [96]. They first discriminated between vocal and non-vocal sections in a mixture by using different combinations of features, such as MFCCs [25], perceptual linear predictive (PLP) coefficients [97], and log frequency power coefficients (LFPC) [98], and training two classifiers, namely neural networks and support vector machines (SVM). They then applied redundancy reduction techniques on the TF representation of the mixture to separate the sources [99], by using NMF (or ICA). The components were then grouped as vocal and non-vocal by reusing a vocal/non-vocal classifier with MFCC, LFPC, and PLP coefficients.

Chanrungutai and Ratanamahatana proposed to use NMF with automatic component selection [100], [101]. They first decomposed the mixture spectrogram using NMF with a fixed number of basis components. They then removed the components with brief rhythmic and long-lasting continuous events, assuming that they correspond to instrumental sounds. They finally used the remaining components to reconstruct the singing voice, after refining them using a high-pass filter.

Marxer and Janer proposed an approach based on a Tikhonov regularization [102] as an alternative to NMF, for singing voice separation [103]. Their method sacrificed the non-negativity constraints of the NMF in exchange for a computationally less expensive solution for spectrum decomposition, making it more interesting for low-latency scenarios.

Yang et al. proposed a Bayesian NMF approach [104], [105]. Following the approaches in [106] and [107], they used a Poisson distribution for the likelihood function and exponential distributions for the model parameters in the NMF algorithm, and derived a variational Bayesian EM algorithm [32] to solve the NMF problem. They also adaptively determined the number of bases from the mixture. They finally grouped the bases into singing voice and background music by using a *k*-means clustering algorithm [108] or an NMF-based clustering algorithm.

In a different manner, Smaragdis and Mysore proposed a user-guided approach for removing sounds from mixtures by humming the target sound to be removed, for example a vocal track [109]. They modeled the mixture using probabilistic latent component analysis (PLCA) [110], another equivalent formulation of NMF. One key feature of exploiting user input was to facilitate the grouping of components into vocals and accompaniment, as humming helped to identify some of the parameters for modeling the vocals.

Nakamuray and Kameoka proposed an $L_p$-norm NMF [111], with $p$ controlling the sparsity of the error. They developed an algorithm for solving this NMF problem based on the auxiliary function principle [112], [113]. Setting an adequate number of bases and $p$ taken as small enough allowed them to estimate the accompaniment as the low-rank decomposition, and the singing voice as the error of the approximation, respectively. Note that, in this case, the singing voice was not explicitly modeled as a sparse component but rather corresponded to the error which happened to be constrained as sparse. The next subsection will actually deal with approaches that explicitly model the vocals as the sparse component.

### B. Low-rank accompaniment, sparse vocals

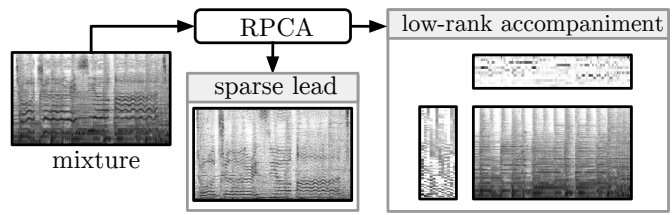

**Fig. 4:** The approaches based on a *low-rank accompaniment, sparse vocals* assumption. As opposed to methods based on NMF, methods based on robust principal component analysis (RPCA) assume the lead signal has a sparse and non-structured spectrogram.

The methods presented in the previous section first compute a decomposition of the mixture into many components that are sorted *a posteriori* as accompaniment or lead. As can be seen, this means they make a low-rank assumption for the accompaniment, but typically *also for the vocals*. However, as can for instance be seen on Figure 1, the spectrogram for the vocals do exhibit much more freedom than accompaniment, and experience shows they are not adequately described by a small number of spectral bases. For this reason, another track of research depicted in Figure 4 focused on using a low-rank assumption on the accompaniment *only*, while assuming the vocals are *sparse and not structured*. This loose assumption means that only a few coefficients from their spectrogram should have significant magnitude, and that they should not feature significant redundancy. Those ideas are in line with robust principal component analysis (RPCA) [114], which is the mathematical tool used by this body of methods, initiated by Huang et al. for singing voice separation [115]. It decomposes a matrix into a sparse and low-rank component.

Sprechmann et al. proposed an approach based on RPCA for online singing voice separation [116]. They used ideas from convex optimization [117], [118] and multi-layer neural networks [119]. They presented two extensions of RPCA and robust NMF models [120]. They then used these extensions in a multi-layer neural network framework which, after an initial training stage, allows online source separation.

Jeong and Lee proposed two extensions of the RPCA model to improve the estimation of vocals and accompaniment from the sparse and low-rank components [121]. Their first extension included the Schatten $p$ and $\ell_p$ norms as generalized nuclear norm optimizations [122]. They also suggested a pre-processing stage based on logarithmic scaling of the mixture TF representation to enhance the RPCA.

Yang also proposed an approach based on RPCA with dictionary learning for recovering low-rank components [123]. He introduced a multiple low-rank representation following the observation that elements of the singing voice can also be recovered by the low-rank component. He first incorporated online dictionary learning methods [124] in his methodology to obtain prior information about the structure of the sources and then incorporated them into the RPCA model.

Chan and Yang then extended RPCA to complex and quaternionic cases with application to singing voice separation [125]. They extended the principal component pursuit (PCP) [114] for solving the RPCA problem by presenting complex and quaternionic proximity operators for the $\ell_1$ and trace-norm regularizations to account for the missing phase information.

### C. Repetitions within the accompaniment

While the rationale behind low-rank methods for lead-accompaniment separation is to exploit the idea that the musical background should be redundant, adopting a low-rank model is not the only way to do it. An alternate way to proceed is to exploit the musical *structure* of songs, to find *repetitions* that can be utilized to perform separation. Just like in RPCA-based methods, the accompaniment is then assumed to be the only source for which repetitions will be found. The unique feature of the methods described here is they combine music structure analysis [126]–[128] with particular ways to exploit the identification of repeated parts of the accompaniment.

Rafii et al. proposed the REpeating Pattern Extraction Technique (REPET) to separate the accompaniment by assuming it is repeating [129]–[131], which is often the case in popular music. This approach, which is representative of this line of research, is represented on Figure 5. First, a repeating period is extracted by a music information retrieval system, such as a beat spectrum [132] in this case. Then, this extracted information is used to estimate the spectrogram of the accompaniment through an averaging of the identified repetitions. From this, a filter is derived.

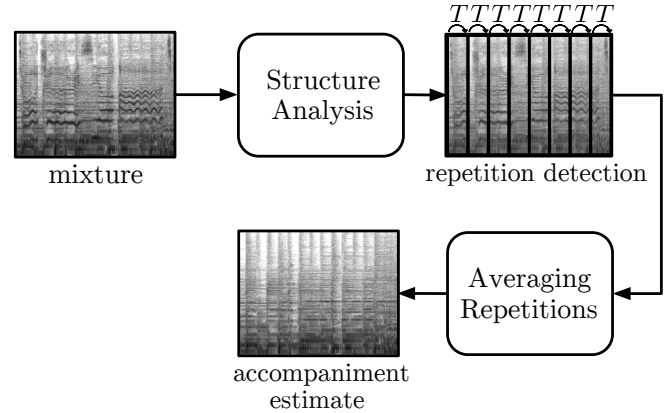

mixture · repetition detection · accompaniment estimate

Fig. 5: The approaches based on a *repetition* assumption for accompaniment. In a first analysis step, repetitions are identified. Then, they are used to build an estimate for the accompaniment spectrogram and proceed to separation.

Seetharaman et al. [133] leveraged the two dimensional Fourier transform (2DFT) of the spectrogram to create an algorithm very similar to REPET. The properties of the 2DFT let them separate the periodic background from the non-periodic vocal melody by deleting peaks in the 2DFT. This eliminated the need to create an explicit model of the periodic audio and without the need to find the period of repetition, both of which are required in REPET.

Liutkus et al. adapted the REPET approach in [129], [130] to handle repeating structures varying along time by modeling the repeating patterns only locally [131], [134]. They first identified a repeating period for every time frame by computing a beat spectrogram as in [132]. Then they estimated the spectrogram of the accompaniment by averaging the time frames in the mixture spectrogram at their local period rate, for every TF bin. From this, they finally extracted the repeating structure by deriving a TF mask.

Rafii et al. further extended the REPET approaches in [129], [130] and [134] to handle repeating structures that are not periodic. To do this, they proposed the REPET-SIM method in [131], [135] to identify repeating frames for every time frame by computing a self-similarity matrix, as in [136]. Then, they estimated the accompaniment spectrogram at every TF bin by averaging the neighbors identified thanks to that similarity matrix. An extension for real-time processing was presented in [137] and a version exploiting user interaction was proposed in [138]. A method close to REPET-SIM was also proposed by FitzGerald in [139].

Liutkus et al. proposed the Kernel Additive modeling (KAM) [140], [141] as a framework which generalizes the REPET approaches in [129]–[131], [134], [135]. They assumed that a source at a TF location can be modeled using its values at other locations through a specified kernel which can account for features such as periodicity, self-similarity, stability over time or frequency, etc. This notably enabled modeling of the accompaniment using more than one repeating pattern. Liutkus et al. also proposed a light version using a fast compression algorithm to make the approach more scalable [142]. The approach was also used for interference reduction in music recordings [143], [144].

With the same idea of exploiting intra-song redundancies for singing voice separation, but through a very different methodology, Moussallam et al. assumed in [145] that all the sources can be decomposed sparsely in the same dictionary and used a matching pursuit greedy algorithm [146] to solve the problem. They integrated the separation process in the algorithm by modifying the atom selection criterion and adding a decision to assign a chosen atom to the repeated source or to the lead signal.

Deif et al. proposed to use multiple median filters to separate vocals from music recordings [147]. They augmented the approach in [148] with diagonal median filters to improve the separation of the vocal component. They also investigated different filter lengths to further improve the separation.

Lee et al. also proposed to use the KAM approach [149]–[152]. They applied the $\beta$-order minimum mean square error (MMSE) estimation [153] to the back-fitting algorithm in KAM to improve the separation. They adaptively calculated a perceptually weighting factor $\alpha$ and the singular value decomposition (SVD)-based factorized spectral amplitude exponent $\beta$ for each kernel component.

### D. Shortcomings

While methods focusing on harmonic models for the lead often fall short in their expressive power for the accompaniment, the methods we reviewed in this section are often observed to suffer exactly from the converse weakness, namely they do not provide an adequate model for the lead signal. Hence, the separated vocals often will feature interference from unpredictable parts from the accompaniment, such as some percussion or effects which occur infrequently.

Furthermore, even if the musical accompaniment will exhibit more redundancy, the vocals part will also be redundant to some extent, which is poorly handled by these methods. When the lead signal is not vocals but played by some lead instrument, its redundancy is even more pronounced, because the notes it plays lie in a reduced set of fundamental frequencies. Consequently, such methods would include the redundant parts of the lead within the accompaniment estimate, for example, a steady humming by a vocalist.

## V. JOINT MODELS FOR LEAD AND ACCOMPANIMENT

In the previous sections, we reviewed two important bodies of literature, focused on modeling either the lead or the accompaniment parts of music recordings, respectively. While each approach showed its own advantages, it also featured its own drawbacks. For this reason, some researchers devised methods combining ideas for modeling both the lead and the accompaniment sources, and thus benefiting from both approaches. We now review this line of research.

### A. Using music structure analysis to drive learning

The first idea we find in the literature is to augment methods for accompaniment modeling with the prior identification of sections where the vocals are present or absent. In the case of the low rank models discussed in Sections IV-A and IV-B, such a strategy indeed dramatically improves performance.

Raj et al. proposed an approach in [154] that is based on the PLCA formulation of NMF [155], and extends their prior work [156]. The parameters for the frequency distribution of the background music are estimated from the background music-only segments, and the rest of the parameters from the singing voice+background music segments, assuming a priori identified vocal regions.

Han and Chen also proposed a similar approach for melody extraction based on PLCA [157], which includes a further estimate of the melody from the vocals signal by an auto-correlation technique similar to [158].

Gómez et al. proposed to separate the singing voice from the guitar accompaniment in flamenco music to help with melody transcription [159]. They first manually segmented the mixture into vocal and non-vocal regions. They then learned percussive and harmonic bases from the non-vocal regions by using an unsupervised NMF percussive/harmonic separation approach [93], [160]. The vocal spectrogram was estimated by keeping the learned percussive and harmonic bases fixed.

Papadopoulos and Ellis proposed a signal-adaptive formulation of RPCA which incorporates music content information to guide the recovery of the sparse and low-rank components [161]. Prior musical knowledge, such as predominant melody, is used to regularize the selection of active coefficients during the optimization procedure.

In a similar manner, Chan et al. proposed to use RPCA with vocal activity information [162]. They modified the RPCA algorithm to constraint parts of the input spectrogram to be non-sparse to account for the non-vocal parts of the singing voice.

A related method was proposed by Jeong and Lee in [163], using RPCA with a weighted $l_1$-norm. They replaced the uniform weighting between the low-rank and sparse components in the RPCA algorithm by an adaptive weighting based on the variance ratio between the singing voice and the accompaniment. One key element of the method is to incorporate vocal activation information in the weighting.

### B. Factorization with a known melody

While using only the knowledge of vocal activity as described above already yields an increase of performance over methods operating blindly, many authors went further to also incorporate the fact that vocals often have a strong melody line. Some redundant model is then assumed for the accompaniment, while also enforcing a harmonic model for the vocals.

An early method to achieve this is depicted in Figure 6 and was proposed by Virtanen et al. in [164]. They estimated the pitch of the vocals in the mixture by using a melody transcription algorithm [63] and derived a binary TF mask to identify where vocals are not present. They then applied NMF on the remaining non-vocal segments to learn a model for the background.

Wang and Ou also proposed an approach which combines melody extraction and NMF-based soft masking [165]. They identified accompaniment, unvoiced, and voiced segments in the mixture using an HMM model with MFCCs and GMMs. They then estimated the pitch of the vocals from the voiced segments using the method in [166] and an HMM with the Viterbi algorithm as in [167]. They finally applied a soft mask to separate voice and accompaniment.

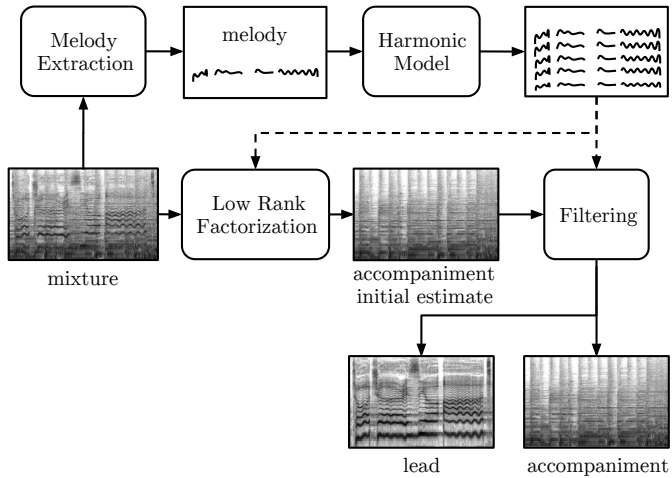

Fig. 6: Factorization informed with the melody. First, melody extraction is performed on the mixture. Then, this information is used to drive the estimation of the accompaniment: TF bins pertaining to the lead should not be taken into account for estimating the accompaniment model.

Rafii et al. investigated the combination of an approach for modeling the background and an approach for modeling the melody [168]. They modeled the background by deriving a rhythmic mask using the REPET-SIM algorithm [135] and the melody by deriving a harmonic mask using a pitch-based algorithm [169]. They proposed a parallel and a sequential combination of those algorithms.

Venkataramani et al. proposed an approach combining sinusoidal modeling and matrix decomposition, which incorporates prior knowledge about singer and phoneme identity [170]. They applied a predominant pitch algorithm on annotated sung regions [171] and performed harmonic sinusoidal modeling [172]. Then, they estimated the spectral envelope of the vocal component from the spectral envelope of the mixture using a phoneme dictionary. After that, a spectral envelope dictionary representing sung vowels from song segments of a given singer was learned using an extension of NMF [173], [174]. They finally estimated a soft mask using the singer-vowel dictionary to refine and extract the vocal component.

Ikemiya et al. proposed to combine RPCA with pitch estimation [175], [176]. They derived a mask using RPCA [115] to separate the mixture spectrogram into singing voice and accompaniment components. They then estimated the fundamental frequency contour from the singing voice component based on [177] and derived a harmonic mask. They integrated the two masks and resynthesized the singing voice and accompaniment signals. Dobashi et al. then proposed to use that singing voice separation approach in a music performance assistance system [178].

Hu and Liu proposed to combine approaches based on matrix decomposition and pitch information for singer identification [179]. They used non-negative matrix partial co-factorization [173], [180] which integrates prior knowledge about the singing voice and the accompaniment, to separate the mixture into singing voice and accompaniment portions. They then identified the singing pitch from the singing voice portions using [181] and derived a harmonic mask as in [182], and finally reconstructed the singing voice using a missing feature method [183]. They also proposed to add temporal and sparsity criteria to their algorithm [184].

That methodology was also adopted by Zhang et al. in [185], that followed the framework of the pitch-based approach in [66], by performing singing voice detection using an HMM classifier, singing pitch detection using the algorithm in [186], and singing voice separation using a binary mask. Additionally, they augmented that approach by analyzing the latent components of the TF matrix using NMF in order to refine the singing voice and accompaniment.

Zhu et al. [187] proposed an approach which is also representative of this body of literature, with the pitch detection algorithm being the one in [181] and binary TF masks used for separation after NMF.

### C. Joint factorization and melody estimation

The methods presented above put together the ideas of modeling the lead (typically the vocals) as featuring a melodic harmonic line and the accompaniment as redundant. As such, they already exhibit significant improvement over approaches only applying one of these ideas as presented in Sections III and IV, respectively. However, these methods above are still restricted in the sense that the analysis performed on each side cannot help improve the other one. In other words, the estimation of the models for the lead and the accompaniment are done sequentially. Another idea is to proceed *jointly*.

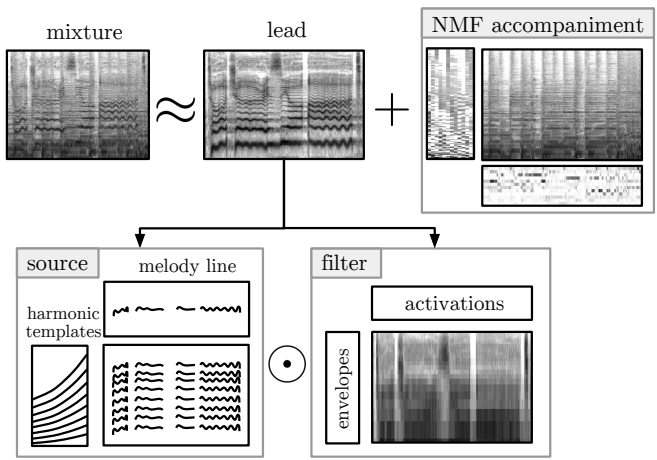

Fig. 7: Joint estimation of the lead and accompaniment, the former one as a source-filter model and the latter one as an NMF model.

A seminal work in this respect was done by Durrieu et al. using a source-filter and NMF model [188]–[190], depicted in Figure 7. Its core idea is to decompose the mixture spectrogram as the sum of two terms. The first term accounts for the lead and is inspired by the source-filter model described in Section II: it is the element-wise product of an *excitation* spectrogram with a *filter* spectrogram. The former one can be understood as harmonic combs activated by the melodic line, while the latter one modulates the envelope and is assumed low-rank because few phonemes are used. The second term accounts for the accompaniment and is modeled with a standard NMF. In [188]–[190], they modeled the lead by using a GMM-based model [191] and a glottal source model [192], and the accompaniment by using an instantaneous mixture model [193] leading to an NMF problem [94]. They jointly estimated the parameters of their models by maximum likelihood estimation using an iterative algorithm inspired by [194] with multiplicative update rules developed in [91]. They also extracted the melody by using an algorithm comparable to the Viterbi algorithm, before re-estimating the parameters and finally performing source separation using Wiener filters [195]. In [196], they proposed to adapt their model for user-guided source separation.

The joint modeling of the lead and accompaniment parts of a music signal was also considered by Fuentes et al. in [197], that introduced the idea of using a log-frequency TF representation called the constant-Q transform (CQT) [198]–[200]. The advantage of such a representation is that a change in pitch corresponds to a simple translation in the TF plane, instead of a scaling as in the STFT. This idea was used along the creation of a user interface to guide the decomposition, in line with what was done in [196].

Joder and Schuller used the source-filter NMF model in [201], additionally exploiting MIDI scores [202]. They synchronized the MIDI scores to the audio using the alignment algorithm in [203]. They proposed to exploit the score information through two types of constraints applied in the model. In a first approach, they only made use of the information regarding whether the leading voice is present or not in each frame. In a second approach, they took advantage of both time and pitch information on the aligned score.

Zhao et al. proposed a score-informed leading voice separation system with a weighting scheme [204]. They extended the system in [202], which is based on the source-filter NMF model in [201], by using a Laplacian or a Gaussian-based mask on the NMF activation matrix to enhance the likelihood of the score-informed pitch candidates.

Jointly estimating accompaniment and lead allowed for some research in correctly estimating the unvoiced parts of the lead, which is the main issue with purely harmonic models, as highlighted in Section III-C. In [201], [205], Durrieu et al. extended their model to account for the unvoiced parts by adding white noise components to the voice model.

In the same direction, Janer and Marxer proposed to separate unvoiced fricative consonants using a semi-supervised NMF [206]. They extended the source-filter NMF model in [201] using a low-latency method with timbre classification to estimate the predominant pitch [87]. They approximated the fricative consonants as an additive wideband component, training a model of NMF bases. They also used the transient quality to differentiate between fricatives and drums, after extracting transient time points using the method in [207].

Similarly, Marxer and Janer then proposed to separately model the singing voice breathiness [208]. They estimated the breathiness component by approximating the voice spectrum as a filtered composition of a glottal excitation and a wide-band component. They modeled the magnitude of the voice spectrum using the model in [209] and the envelope of the voice excitation using the model in [192]. They estimated the pitch using the method in [87]. This was all integrated into the source-filter NMF model.

The body of research initiated by Durrieu et al. in [188] consists of using algebraic models more sophisticated than one simple matrix product, but rather inspired by musicological knowledge. Ozerov et al. formalized this idea through a general framework and showed its application for singing voice separation [210]–[212].

Finally, Hennequin and Rigaud augmented their model to account for long-term reverberation, with application to singing voice separation [213]. They extended the model in [214] which allows extraction of the reverberation of a specific source with its dry signal. They combined this model with the source-filter NMF model in [189].

### D. Different constraints for different sources

Algebraic methods that decompose the mixture spectrogram as the sum of the lead and accompaniment spectrograms are based on the minimization of a *cost* or *loss function* which measures the error between the approximation and the observation. While the methods presented above for lead and accompaniment separation did propose more sophisticated models with parameters explicitly pertaining to the lead or the accompaniment, another option that is also popular in the dedicated literature is to modify the cost function of an optimization algorithm for an existing algorithm (e.g., RPCA), so that one part of the resulting components would preferentially account for one source or another.

This approach can be exemplified by the harmonic-percussive source separation method (HPSS), presented in [160], [215], [216]. It consists in filtering a mixture spectrogram so that horizontal lines go in a so-called *harmonic* source, while its vertical lines go into a *percussive* source. Separation is then done with TF masking. Of course, such a method is not adequate for lead and accompaniment separation *per se*, because all the harmonic content of the accompaniment is classified as harmonic. However, it shows that *nonparametric* approaches are also an option, provided the cost function itself is well chosen for each source.

This idea was followed by Yang in [217] who proposed an approach based on RPCA with the incorporation of harmonicity priors and a back-end drum removal procedure to improve the decomposition. He added a regularization term in the algorithm to account for harmonic sounds in the low-rank component and used an NMF-based model trained for drum separation [211] to eliminate percussive sounds in the sparse component.

Jeong and Lee proposed to separate a vocal signal from a music signal [218], extending the HPSS approach in [160],

[215]. Assuming that the spectrogram of the signal can be represented as the sum of harmonic, percussive, and vocal components, they derived an objective function which enforces the temporal and spectral continuity of the harmonic and percussive components, respectively, similarly to [160], but also the sparsity of the vocal component. Assuming non-negativity of the components, they then derived iterative update rules to minimize the objective function. Ochiai et al. extended this work in [219], notably by imposing harmonic constraints for the lead.

Watanabe et al. extended RPCA for singing voice separation [220]. They added a harmonicity constraint in the objective function to account for harmonic structures, such as in vocal signals, and regularization terms to enforce the non-negativity of the solution. They used the generalized forward-backward splitting algorithm [221] to solve the optimization problem. They also applied post-processing to remove the low frequencies in the vocal spectrogram and built a TF mask to remove time frames with low energy.

Going beyond smoothness and harmonicity, Hayashi et al. proposed an NMF with a constraint to help separate periodic components, such as a repeating accompaniment [222]. They defined a periodicity constraint which they incorporated in the objective function of the NMF algorithm to enforce the periodicity of the bases.

### E. Cascaded and iterated methods

In their effort to propose separation methods for the lead and accompaniment in music, some authors discovered that very different methods often have complementary strengths. This motivated the *combination* of methods. In practice, there are several ways to follow this line of research.

One potential route to achieve better separation is to *cascade* several methods. This is what FitzGerald and Gainza proposed in [216] with multiple median filters [148]. They used a median-filter based HPSS approach at different frequency resolutions to separate a mixture into harmonic, percussive, and vocal components. They also investigated the use of STFT or CQT as the TF representation and proposed a post-processing step to improve the separation with tensor factorization techniques [223] and non-negative partial co-factorization [180].

The two-stage HPSS system proposed by Tachibana et al. in [224] proceeds the same way. It is an extension of the melody extraction approach in [225] and was applied for karaoke in [226]. It consists in using the optimization-based HPSS algorithm from [160], [215], [227], [228] at different frequency resolutions to separate the mixture into harmonic, percussive, and vocal components.

HPSS was not the only separation module considered as the building block of combined lead and accompaniment separation approaches. Deif et al. also proposed a multi-stage NMF-based algorithm [229], based on the approach in [230]. They used a local spectral discontinuity measure to refine the non-pitched components obtained from the factorization of the long window spectrogram and a local temporal discontinuity measure to refine the non-percussive components obtained from factorization of the short window spectrogram.

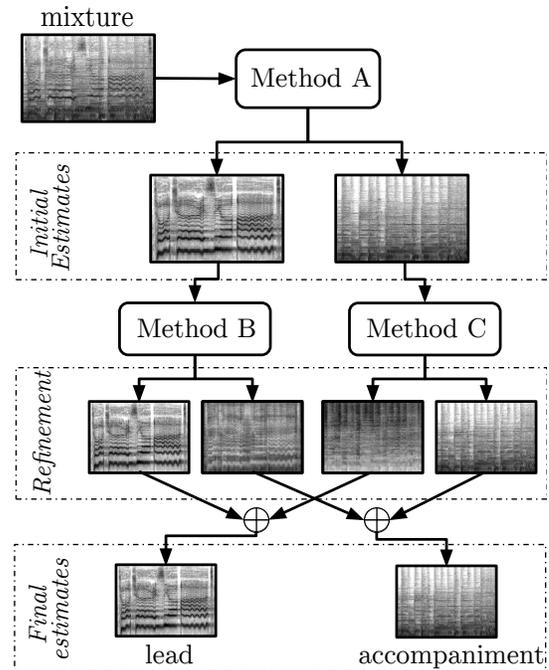

Fig. 8: Cascading source separation methods. The results from method A is improved by applying methods B and C on its output, which are specialized in reducing interferences from undesired sources in each signal.

Finally, this cascading concept was considered again by Driedger and Müller in [231], that introduces a processing pipeline for the outputs of different methods [115], [164], [232], [233] to obtain an improved separation quality. Their core idea is depicted in Figure 8 and combines the output of different methods in a specific order to improve separation.

Another approach for improving the quality of separation when using several separation procedures is not to restrict the number of such iterations from one method to another, but rather to iterate them many times until satisfactory results are obtained. This is what is proposed in Hsu et al. in [234], extending the algorithm in [235]. They first estimated the pitch range of the singing voice by using the HPSS method in [160], [225]. They separated the voice given the estimated pitch using a binary mask obtained by training a multilayer perceptron [236] and re-estimated the pitch given the separated voice. Voice separation and pitch estimation are then iterated until convergence.

As another iterative method, Zhu et al. proposed a multi-stage NMF [230], using harmonic and percussive separation at different frequency resolutions similar to [225] and [216]. The main originality of their contribution was to iterate the refinements instead of applying it only once.

An issue with such iterated methods lies in how to decide whether convergence is obtained, and it is not clear whether the quality of the separated signals will necessarily improve. For this reason, Bryan and Mysore proposed a user-guided approach based on PLCA, which can be applied for the separation of the vocals [237]–[239]. They allowed a user to make annotations on the spectrogram of a mixture, incorporated the feedback as constraints in a PLCA model [110], [156], and used a posterior regularization technique [240] to refine the estimates, repeating the process until the user is satisfied with the results. This is similar to the way Ozerov et al. proposed to take user input into account in [241].

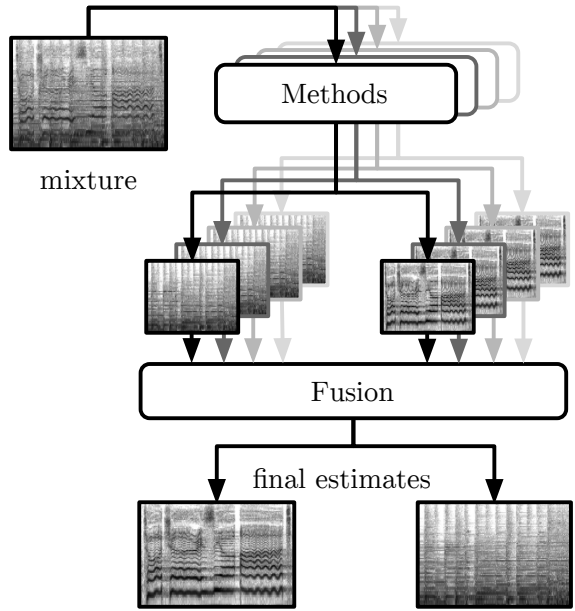

Fig. 9: Fusion of separation methods. The output of many separation methods is fed into a fusion system that combines them to produce a single estimate.

A principled way to aggregate the result of many source separation systems to obtain one single estimate that is consistently better than all of them was presented by Jaureguiberry et al. in their *fusion framework*, depicted in Figure 9. It takes advantage of multiple existing approaches, and demonstrated its application to singing voice separation [242]–[244]. They investigated fusion methods based on non-linear optimization, Bayesian model averaging [245], and deep neural networks (DNN).

As another attempt to design an efficient fusion method, McVicar et al. proposed in [246] to combine the outputs of RPCA [115], HPSS [216], Gabor filtered spectrograms [247], REPET [130] and an approach based on deep learning [248]. To do this, they used different classification techniques to build the aggregated TF mask, such as a logistic regression model or a conditional random field (CRF) trained using the method in [249] with time and/or frequency dependencies.

Manilow et al. trained a neural network to predict quality of source separation for three source separation algorithms, each leveraging a different cue - repetition, spatialization, and harmonicity/pitch proximity [250]. The method estimates separation quality of the lead vocals for each algorithm, using only the original audio mixture and separated source output. These estimates were used to guide switching between algorithms along time.

## F. Source-dependent representations

In the previous section, we stated that some authors considered iterating separation at different frequency resolutions, i.e., using different TF representations [216], [224], [229]. This can be seen as a combination of different methods. However, this can also be seen from another perspective as based on picking specific *representations*.

Wolf et al. proposed an approach using rigid motion segmentation, with application to singing voice separation [251], [252]. They introduced harmonic template models with amplitude and pitch modulations defined by a velocity vector. They applied a wavelet transform [253] on the harmonic template models to build an audio image where the amplitude and pitch dynamics can be separated through the velocity vector. They then derived a velocity equation, similar to the optical flow velocity equation used in images [254], to segment velocity components. Finally, they identified the harmonic templates which model different sources in the mixture and separated them by approximating the velocity field over the corresponding harmonic template models.

Yen et al. proposed an approach using spectro-temporal modulation features [255], [256]. They decomposed a mixture using a two-stage auditory model which consists of a cochlear module [257] and cortical module [258]. They then extracted spectro-temporal modulation features from the TF units and clustered the TF units into harmonic, percussive, and vocal components using the EM algorithm and resynthesized the estimated signals.

Chan and Yang proposed an approach using an informed group sparse representation [259]. They introduced a representation built using a learned dictionary based on a chord sequence which exhibits group sparsity [260] and which can incorporate melody annotations. They derived a formulation of the problem in a manner similar to RPCA and solved it using the alternating direction method of multipliers [261]. They also showed a relation between their representation and the low-rank representation in [123], [262].

## G. Shortcomings

The large body of literature we reviewed in the preceding sections is concentrated on choosing adequate models for the lead and accompaniment parts of music signals in order to devise effective signal processing methods to achieve separation. From a higher perspective, their common feature is to guide the separation process in a *model-based way*: first, the scientist has some idea regarding characteristics of the lead signal and/or the accompaniment, and then an algorithm is designed to exploit this knowledge for separation.

Model-based methods for lead and accompaniment separation are faced with a common risk that their core assumptions will be violated for the signal under study. For instance, the lead to be separated may not be harmonic but saturated vocals or the accompaniment may not be repetitive or redundant, but rather always changing. In such cases, model-based methods are prone to large errors and poor performance.

## VI. Data-driven approaches

A way to address the potential caveats of model-based separation behaving badly in case of violated assumptions is to avoid making assumptions altogether, but rather to let the model be learned from a large and representative database of examples. This line of research leads to *data-driven* methods, for which researchers are concerned about directly estimating a mapping between the mixture and either the TF mask for separating the sources, or their spectrograms to be used for designing a filter.

As may be foreseen, this strategy based on machine learning comes with several challenges of its own. First, it requires considerable amounts of data. Second, it typically requires a high-capacity learner (many tunable parameters) that can be prone to over-fitting the training data and therefore not working well on the audio it faces when deployed.

### A. Datasets

Building a good data-driven method for source separation relies heavily on a training dataset to learn the separation model. In our case, this not only means obtaining a set of musical songs, but also their constitutive accompaniment and lead sources, summing up to the mixtures. For professionally-produced or recorded music, the separated sources are often either unavailable or private. Indeed, they are considered amongst the most precious assets of right holders, and it is very difficult to find isolated vocals and accompaniment of professional bands that are freely available for the research community to work on without copyright infringements.

Another difficulty arises when considering that the different sources in a musical content do share some common orchestration and are not superimposed in a random way, prohibiting simply summing isolated random notes from instrumental databases to produce mixtures. This contrasts with the speech community which routinely generates mixtures by summing noise data [263] and clean speech [264].

Furthermore, the temporal structures in music signals typically spread over long periods of time and can be exploited to achieve better separation. Additionally, short excerpts do not often comprise parts where the lead signal is absent, although a method should learn to deal with that situation. This all suggests that including full songs in the training data is preferable over short excerpts.

Finally, professional recordings typically undergo sophisticated sound processing where panning, reverberation, and other sound effects are applied to each source separately, and also to the mixture. To date, simulated data sets have poorly mimicked these effects [265]. Many separation methods make assumptions about the mixing model of the sources, e.g., assuming it is linear (i.e., does not comprise effects such as dynamic range compression). It is quite common that methods giving extremely good performance for linear mixtures completely break down when processing published musical recordings. Training and test data should thus feature realistic audio engineering to be useful for actual applications.

In this context, the development of datasets for lead and accompaniment separation was a long process. In early times, it was common for researchers to test their methods on some private data. To the best of our knowledge, the first attempt at releasing a public dataset for evaluating vocals and accompaniment separation was the Music Audio Signal Separation (MASS) dataset [266]. It strongly boosted research in the area, even if it only featured 2.5 minutes of data. The breakthrough was made possible by some artists who made their mixed-down audio, as well as its constitutive stems (unmixed tracks), available under open licenses such as Creative Commons, or authorized scientists to use their material for research.

The MASS dataset then formed the core content of the early Signal Separation Evaluation Campaigns (SiSEC) [267], which evaluate the quality of various music separation methods [268]–[272]. SiSEC always had a strong focus on vocals and accompaniment separation. For a long time, vocals separation methods were very demanding computationally and it was already considered extremely challenging to separate excerpts of only a few seconds.

In the following years, new datasets were proposed that improved over the MASS dataset in many directions. We briefly describe the most important ones, summarized in Table I.

- The QUASI dataset was proposed to study the impact of different mixing scenarios on the separation quality. It consists of the same tracks as in the MASS dataset, but kept full length and mixed by professional sound engineers.
- The MIR-1K and iKala datasets were the first attempts to scale vocals separation up. They feature a higher number of samples than the previously available datasets. However, they consist of mono signals of very short and amateur karaoke recordings.
- The ccMixter dataset was proposed as the first dataset to feature many full-length stereo tracks. Each one comes with a vocals and an accompaniment source. Although it is stereo, it often suffers from simplistic mixing of sources, making it unrealistic in some aspects.
- MedleyDB has been developed as a dataset to serve many purposes in music information retrieval. It consists of more than 100 full-length recordings, with all their constitutive sources. It is the first dataset to provide such a large amount of data to be used for audio separation research (more than 7 hours). Among all the material present in that dataset, 63 tracks feature singing voice.
- DSD100 was presented for SiSEC 2016. It features 100 full-length tracks originating from the 'Mixing Secret' Free Multitrack Download Library[1] of the Cambridge Music Technology, which is freely usable for research and educational purposes.

Finally, we present here the MUSDB18 dataset, putting together tracks from MedleyDB, DSD100, and other new musical material. It features 150 full-length tracks, and has been constructed by the authors of this paper so as to address all the limitations we identified above:

- It only features full-length tracks, so that the handling of long-term musical structures, and of silent regions in the lead/vocal signal, can be evaluated.

---

[1] http://www.cambridge-mt.com/ms-mtk.htm

- It only features stereo signals which were mixed using professional digital audio workstations. This results in quality stereo mixes which are representative of real application scenarios.
- As with DSD100, a design choice of MUSDB18 was to split the signals into 4 predefined categories: bass, drums, vocals, and other. This contrasts with the enhanced granularity of MedleyDB that offers more types of sources, but it strongly promotes automation of the algorithms.
- Many musical genres are represented in MUSDB18, for example, jazz, electro, metal, etc.
- It is split into a development (100 tracks, 6.5 h) and a test dataset (50 tracks, 3.5 h), for the design of data-driven separation methods.

All details about this freely available dataset and its accompanying software tools may be found in its dedicated website[2].

In any case, it can be seen that datasets of sufficient duration to build data-driven separation methods were only created recently.

### B. Algebraic approaches

A natural way to exploit a training database was to learn some parts of the model to guide the estimation process into better solutions. Work on this topic may be traced back to the suggestion of Ozerov et al. in [276] to learn spectral template models based on a database of isolated sources, and then to adapt this dictionary of templates on the mixture using the method in [277].

The exploitation of training data was formalized by Smaragdis et al. in [110] in the context of source separation within the supervised and semi-supervised PLCA framework. The core idea of this probabilistic formulation, equivalent to NMF, is to learn some spectral bases from the training set which are then kept fixed at separation time.

In the same line, Ozerov et al. proposed an approach using Bayesian models [191]. They first segmented a song into vocal and non-vocal parts using GMMs with MFCCs. Then, they adapted a general music model on the non-vocal parts of a particular song by using the maximum a posteriori (MAP) adaptation approach in [278]

Ozerov et al. later proposed a framework for source separation which generalizes several approaches given prior information about the problem and showed its application for singing voice separation [210]–[212]. They chose the local Gaussian model [279] as the core of the framework and allowed the prior knowledge about each source and its mixing characteristics using user-specified constraints. Estimation was performed through a generalized EM algorithm [32].

Rafii et al. proposed in [280] to address the main drawback of the repetition-based methods described in Section IV-C, which is the weakness of the model for vocals. For this purpose, they combined the REPET-SIM model [135] for the accompaniment with a NMF-based model for singing voice learned from a voice dataset.

As yet another example of using training data for NMF, Boulanger-Lewandowski et al. proposed in [281] to exploit long-term temporal dependencies in NMF, embodied using recurrent neural networks (RNN) [236]. They incorporated RNN regularization into the NMF framework to temporally constrain the activity matrix during the decomposition, which can be seen as a generalization of the non-negative HMM in [282]. Furthermore, they used supervised and semi-supervised NMF algorithms on isolated sources to train the models, as in [110].

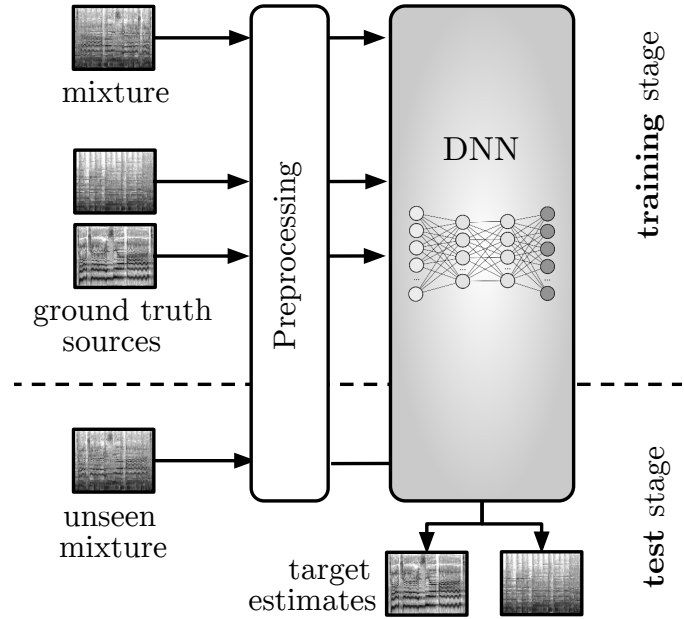

Fig. 10: General architecture for methods exploiting deep learning. The network inputs the mixture and outputs either the sources spectrograms or a TF mask. Methods usually differ in their choice for a network architecture and the way it is learned using the training data.

### C. Deep neural networks

Taking advantage of the recent availability of sufficiently large databases of isolated vocals along with their accompaniment, several researchers investigated the use of machine learning methods to directly estimate a mapping between the mixture and the sources. Although end-to-end systems inputting and outputting the waveforms have already been proposed in the speech community [283], they are not yet available for music source separation. This may be due to the relative small size of music separation databases, at most 10 h today. Instead, most systems feature pre and post-processing steps that consist in computing classical TF representations and building TF masks, respectively. Although such end-to-end systems will inevitably be proposed in the near future, the common structure of deep learning methods for lead and accompaniment separation usually corresponds for now to the one depicted in Figure 10. From a general perspective, we may say that most current methods mainly differ in the structure picked for the network, as well as in the way it is learned.

Providing a thorough introduction to deep neural networks is out of the scope of this paper. For our purpose, it suffices to mention that they consist of a cascade of several possibly non-linear transformations of the input, which are learned during a training stage. They are shown to effectively learn representations and mappings, provided enough data is available for estimating their parameters [284]–[286]. Different architectures for neural networks may be combined/cascaded together, and many architectures were proposed in the past, such as feedforward fully-connected neural networks (FNN), convolutional neural networks (CNN), or RNN and variants such as the long short-term memory (LSTM) and the gated-recurrent units (GRU). Training of such functions is achieved by stochastic gradient descent [287] and associated algorithms, such as backpropagation [288] or backpropagation through time [236] for the case of RNNs.

TABLE I: Summary of datasets available for lead and accompaniment separation. Tracks without vocals were omitted in the statistics.

| Dataset | Year | Reference(s) | URL | Tracks | Track duration (s) | Full/stereo? |
|---------|------|--------------|-----|--------|--------------------|--------------|
| MASS | 2008 | [266] | http://www.mtg.upf.edu/download/datasets/mass | 9 | $16 \pm 7$ | no / yes |
| MIR-1K | 2010 | [74] | https://sites.google.com/site/unvoicedsoundseparation/mir-1k | 1,000 | $8 \pm 8$ | no / no |
| QUASI | 2011 | [270], [273] | http://www.tsi.telecom-paristech.fr/aao/en/2012/03/12/quasi/ | 5 | $206 \pm 21$ | yes / yes |
| ccMixter | 2014 | [141] | http://www.loria.fr/~aliutkus/kam/ | 50 | $231 \pm 77$ | yes / yes |
| MedleyDB | 2014 | [274] | http://medleydb.weebly.com/ | 63 | $206 \pm 121$ | yes / yes |
| iKala | 2015 | [162] | http://mac.citi.sinica.edu.tw/ikala/ | 206 | 30 | no / no |
| DSD100 | 2015 | [271] | sisec17.audiolabs-erlangen.de | 100 | $251 \pm 60$ | yes / yes |
| MUSDB18 | 2017 | [275] | https://sigsep.github.io/musdb | 150 | $236 \pm 95$ | yes / yes |

To the best of our knowledge, Huang et al. were the first to propose deep neural networks, RNNs here [289], [290], for singing voice separation in [248], [291]. They adapted their framework from [292] to model all sources simultaneously through masking. Input and target functions were the mixture magnitude and a joint representation of the individual sources. The objective was to estimate jointly either singing voice and accompaniment music, or speech and background noise from the corresponding mixtures.

Modeling the temporal structures of both the lead and the accompaniment is a considerable challenge, even when using DNN methods. As an alternative to the RNN approach proposed by Huang et al. in [248], Uhlich et al. proposed the usage of FNNs [293] whose input consists of *supervectors* of a few consecutive frames from the mixture spectrogram. Later in [294], the same authors considered the use of bi-directional LSTMs for the same task.

In an effort to make the resulting system less computationally demanding at separation time but still incorporating dynamic modeling of audio, Simpson et al. proposed in [295] to predict binary TF masks using deep CNNs, which typically utilize fewer parameters than the FNNs. Similarly, Schlueter proposed a method trained to detect singing voice using CNNs [296]. In that case, the trained network was used to compute *saliency maps* from which TF masks can be computed for singing voice separation. Chandna et al. also considered CNNs for lead separation in [297], with a particular focus on low-latency.

The classical FNN, LSTM and CNN structures above served as baseline structures over which some others tried to improve. As a first example, Mimilakis et al. proposed to use a hybrid structure of FNNs with skip connections to separate the lead instrument for purposes of remixing jazz recordings [298]. Such skip connections allow to propagate the input spectrogram to intermediate representations within the network, and mask it similarly to the operation of TF masks. As advocated, this enforces the networks to approximate a TF masking process. Extensions to temporal data for singing voice separation were presented in [299], [300]. Similarly, Jansson et al. proposed to propagate the spectral information computed by convolutional layers to intermediate representations [301]. This propagation aggregates intermediate outputs to proceeding layer(s). The output of the last layer is responsible for masking the input mixture spectrogram. In the same vein, Takahashi et al. proposed to use skip connections via element-wise addition through representations computed by CNNs [302].

Apart from the structure of the network, the way it is trained, comprising how the targets are computed, has a tremendous impact on performance. As we saw, most methods operate on defining TF masks or estimating magnitude spectrograms. However, other methods were proposed based on deep clustering [303], [304], where TF mask estimation is seen as a clustering problem. Luo et al. investigated both approaches in [305] by proposing deep bidirectional LSTM networks capable of outputting both TF masks or features to use as in deep clustering. Kim and Smaragdis proposed in [306] another way to learn the model, in a denoising auto-encoding fashion [307], again utilizing short segments of the mixture spectrogram as an input to the network, as in [293].

As the best network structure may vary from one track to another, some authors considered a fusion of methods, in a manner similar to the method [242] presented above. Grais et. al [308], [309] proposed to aggregate the results from an ensemble of feedforward DNNs to predict TF masks for separation. An improvement was presented in [310], [311] where the inputs to the fusion network were separated signals, instead of TF masks, aiming at enhancing the reconstruction of the separated sources.

As can be seen the use of deep learning methods for the design of lead and accompaniment separation has already stimulated a lot of research, although it is still in its infancy. Interestingly, we also note that using audio and music specific knowledge appears to be fundamental in designing effective systems. As an example of this, the contribution from Nie et al. in [312] was to include the construction of the TF mask as an extra non-linearity included in a recurrent network. This is an exemplar of where signal processing elements, such as filtering through masking, are incorporated as a building block of the machine learning method.

The network structure is not the only thing that can benefit from audio knowledge for better separation. The design of appropriate features is another. While we saw that supervectors of spectrogram patches offered the ability to effectively model time-context information in FNNs [293], Sebastian and Murthy [313] proposed the use of the modified group delay feature representation [314] in their deep RNN architecture. They applied their approach for both singing voice and vocal-violin separation.

Finally, as with other methods, DNN-based separation techniques can also be combined with others to yield improved performance. As an example, Fan et al. proposed to use DNNs to separate the singing voice and to also exploit vocal pitch estimation [315]. They first extracted the singing voice using feedforward DNNs with sigmoid activation functions. They then estimated the vocal pitch from the extracted singing voice using dynamic programming.

### D. Shortcomings

Data-driven methods are nowadays the topic of important research efforts, particularly those based on DNNs. This is notably due to their impressive performance in terms of separation quality, as can, for instance, be noticed below in Section VIII. However, they also come with some limitations.

First, we highlighted that lead and accompaniment separation in music has the very specific problem of scarce data. Since it is very hard to gather large amounts of training data for that application, it is hard to fully exploit learning methods that require large training sets. This raises very specific challenges in terms of machine learning.

Second, the lack of interpretability of model parameters is often mentioned as a significant shortcoming when it comes to applications. Indeed, music engineering systems are characterized by a strong importance of human-computer interactions, because they are used in an artistic context that may require specific needs or results. As of today, it is unclear how to provide user interaction for controlling the millions of parameters of DNN-based systems.

## VII. Including multichannel information

In describing the above methods, we have not discussed the fact that music signals are typically stereophonic. On the contrary, the bulk of methods we discussed focused on designing good spectrogram models for the purpose of filtering mixtures that may be *monophonic*. Such a strategy is called *single-channel* source separation and is usually presented as more challenging than multichannel source separation. Indeed, only TF structure may then be used to discriminate the accompaniment from the lead. In stereo recordings, one further so-called *spatial* dimension is introduced, which is sometimes referred to as *pan*, that corresponds to the perceived *position* of a source in the stereo field. Devising methods to exploit this spatial diversity for source separation has also been the topic of an important body of research that we review now.

### A. Extracting the lead based on panning

In the case of popular music signals, a fact of paramount practical importance is that the lead signal — such as vocals — is very often mixed *in the center*, which means that its energy is approximately the same in left and right channels. On the contrary, other instruments are often mixed at positions to the left or right of the stereo field.

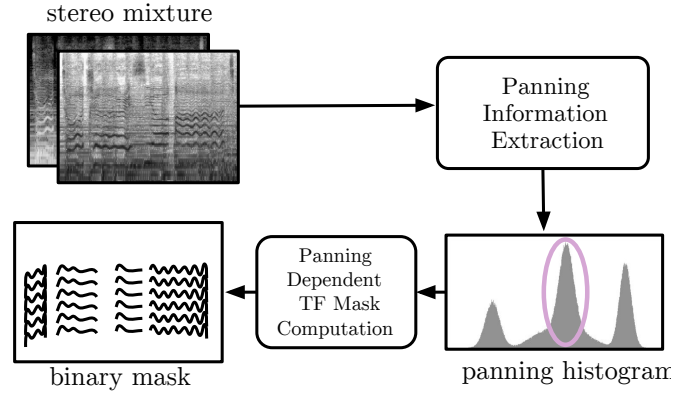

Fig. 11: Separation of the lead based on panning information. A stereo cue called panning allows to design a TF mask.

The general structure of methods extracting the lead based on stereo cues is displayed on Figure 11, introduced by Avendano, who proposed to separate sources in stereo mixtures by using a panning index [316]. He derived a two-dimensional map by comparing left and right channels in the TF domain to identify the different sources based on their panning position [317]. The same methodology was considered by Barry et al. in [318] in his Azimuth Discrimination and Resynthesis (ADRess) approach, with panning indexes computed with differences instead of ratios.

Vinyes et al. also proposed to unmix commercially produced music recordings thanks to stereo cues [319]. They designed an interface similar to [318] where a user can set some parameters to generate different TF filters in real time. They showed applications for extracting various instruments, including vocals.

Cobos and López proposed to separate sources in stereo mixtures by using TF masking and multilevel thresholding [320]. They based their approach on the Degenerate Unmixing Estimation Technique (DUET) [321]. They first derived histograms by measuring the amplitude relationship between TF points in left and right channels. Then, they obtained several thresholds using the multilevel extension of Otsu's method [322]. Finally, TF points were assigned to their related sources to produce TF masks.

Sofianos et al. proposed to separate the singing voice from a stereo mixture using ICA [323]–[325]. They assumed that most commercial songs have the vocals panned to the center and that they dominate the other sources in amplitude. In [323], they proposed to combine a modified version of ADRess with ICA to filter out the other instruments. In [324], they proposed a modified version without ADRess.

Kim et al. proposed to separate centered singing voice in stereo music by exploiting binaural cues, such as inter-channel level and inter-channel phase difference [326]. To this end, they build the pan-based TF mask through an EM algorithm, exploiting a GMM model on these cues.

### B. Augmenting models with stereo

As with using only a harmonic model for the lead signal, using stereo cues in isolation is not always sufficient for good separation, as there can often be multiple sources at the same spatial location. Combining stereo cues with other methods improves performance in these cases.

Cobos and López proposed to extract singing voice by combining panning information and pitch tracking [327]. They first obtained an estimate for the lead thanks to a pan-based method such as [316], and refined the singing voice by using a TF binary mask based on comb-filtering method as in Section III-B. The same combination was proposed by Marxer et al. in [87] in a low-latency context, with different methods used for the binaural cues and pitch tracking blocks.

FitzGerald proposed to combine approaches based on repetition and panning to extract stereo vocals [328]. He first used his nearest neighbors median filtering algorithm [139] to separate vocals and accompaniment from a stereo mixture. He then used the ADRess algorithm [318] and a high-pass filter to refine the vocals and improve the accompaniment. In a somewhat different manner, FitzGerald and Jaiswal also proposed to combine approaches based on repetition and panning to improve stereo accompaniment recovery [329]. They presented an audio inpainting scheme [330] based on the nearest neighbors and median filtering algorithm [139] to recover TF regions of the accompaniment assigned to the vocals after using a source separation algorithm based on panning information.

In a more theoretically grounded manner, several methods based on a probabilistic model were generalized to the multichannel case. For instance, Durrieu et al. extended their source-filter model in [201], [205] to handle stereo signals, by incorporating the panning coefficients as model parameters to be estimated.

Ozerov and Févotte proposed a multichannel NMF framework with application to source separation, including vocals and music [331], [332]. They adopted a statistical model where each source is represented as a sum of Gaussian components [193], and where maximum likelihood estimation of the parameters is equivalent to NMF with the Itakura-Saito divergence [94]. They proposed two methods for estimating the parameters of their model, one that maximized the likelihood of the multichannel data using EM, and one that maximized the sum of the likelihoods of all channels using a multiplicative update algorithm inspired by NMF [90].

Ozerov et al. then proposed a multichannel non-negative tensor factorization (NTF) model with application to user-guided source separation [333]. They modeled the sources jointly by a 3-valence tensor (time/frequency/source) as in [334] which extends the multichannel NMF model in [332]. They used a generalized EM algorithm based on multiplicative updates [335] to minimize the objective function. They incorporated information about the temporal segmentation of the tracks and the number of components per track. Ozerov et al. later proposed weighted variants of NMF and NTF with application to user-guided source separation, including separation of vocals and music [241], [336].

Sawada et al. also proposed multichannel extensions of NMF, tested for separating stereo mixtures of multiple sources, including vocals and accompaniment [337]–[339]. They first defined multichannel extensions of the cost function, namely, Euclidean distance and Itakura-Saito divergence, and derived multiplicative update rules accordingly. They then proposed two techniques for clustering the bases, one built into the NMF model and one performing sequential pair-wise merges.

Finally, multichannel information was also used with DNN models. Nugraha et al. addressed the problem of multichannel source separation for speech enhancement [340], [341] and music separation [342], [343]. In this framework, DNNs are still used for the spectrograms, while more classical EM algorithms [344], [345] are used for estimating the spatial parameters.

### C. Shortcomings

When compared to simply processing the different channels independently, incorporating spatial information in the separation method often comes at the cost of additional computational complexity. The resulting methods are indeed usually more demanding in terms of computing power, because they involve the design of beamforming filters and inversion of covariance matrices. While this is not really an issue for stereophonic music, this may become prohibiting in configurations with higher numbers of channels.

## VIII. EVALUATION

### A. Background

The problem of evaluating the quality of audio signals is a research topic of its own, which is deeply connected to psychoacoustics [346] and has many applications in engineering because it provides an objective function to optimize when designing processing methods. While mean squared error (MSE) is often used for mathematical convenience whenever an error is to be computed, it is a very established fact that MSE is not representative of audio perception [347], [348]. For example, inaudible phase shifts would dramatically increase the MSE. Moreover, it should be acknowledged that the concept of quality is rather application-dependent.

In the case of signal separation or enhancement, processing is often only a part of a whole architecture and a relevant methodology for evaluation is to study the positive or negative impact of this module on the overall performance of the system, rather than to consider it independently from the rest. For example, when embedded in an automatic speech recognition (ASR) system, performance of speech denoising can be assessed by checking whether it decreases word error rate [349].

When it comes to music processing, and more particularly to lead and accompaniment separation, the evaluation of separation quality has traditionally been inspired by work in the audio coding community [347], [350] in the sense that it aims at comparing ground truth vocals and accompaniment with their estimates, just like audio coding compares the original with the compressed signal.

### B. Metrics

As noted previously, MSE-based error measures are not perceptually relevant. For this reason, a natural approach is to have humans do the comparison. The gold-standard for human perceptual studies is the MUlti Stimulus test with Hidden Reference and Anchor (MUSHRA) methodology, that is commonly used for evaluating audio coding [350].

However, it quickly became clear that the specific evaluation of separation quality cannot easily be reduced to a single number, even when achieved through actual perceptual campaigns, but that quality rather depends on the application considered. For instance, karaoke or vocal extraction come with opposing trade-offs between isolation and distortion. For this reason, it has been standard practice to provide different and complementary metrics for evaluating separation that measure the amount of distortion, artifacts, and interference in the results.

While human-based perceptual evaluation is definitely the best way to assess separation quality [351], [352], having computable objective metrics is desirable for several reasons. First, it allows researchers to evaluate performance without setting up costly and lengthy perceptual evaluation campaigns. Second, it permits large-scale training for the fine-tuning of parameters. In this respect, the Blind Source Separation Evaluation (BSS Eval) toolbox [353], [354] provides quality metrics in decibel to account for distortion (SDR), artifacts (SAR), and interferences (SIR). Since it was made available quite early and provides somewhat reasonable correlation with human perception in certain cases [355], [356] it is still widely used to this day.

Even if BSS Eval was considered sufficient for evaluation purposes for a long time, it is based on squared error criteria. Following early work in the area [357], the Perceptual Evaluation of Audio Source Separation (PEASS) toolkit [358]–[360] was introduced as a way to predict perceptual ratings. While the methodology is very relevant, PEASS however was not widely accepted in practice. We believe this is for two reasons. First, the proposed implementation is quite computationally demanding. Second, the perceptual scores it was designed with are more related to speech separation than to music.

Improving perceptual evaluation often requires a large amount of experiments, which is both costly and requires many expert listeners. One way to increase the number of participants is to conduct web-based experiments. In [361], the authors report they were able to aggregate 530 participants in only 8.2 hours and obtained perceptual evaluation scores comparable to those estimated in the controlled lab environment.

Finally, we highlight here that the development of new perceptually relevant objective metrics for singing voice separation evaluation remains an open issue [362]. It is also a highly crucial one for future research in the domain.

### C. Performance (SiSEC 2016)

In this section, we will discuss the performance of 23 source separation methods evaluated on the DSD100, as part of the task for separating professionally-produced music recordings at SiSEC 2016. The methods are listed in Table II, along with the acronyms we use for them, their main references, a very brief summary, and a link to the section where they are described in the text. To date, this stands as the largest evaluation campaign ever achieved on lead and accompaniment separation. The results we discuss here are a more detailed report for SiSEC 2016 [272], presented in line with the taxonomy proposed in this paper.

TABLE II: Methods evaluated during SiSEC 2016.

| Acronym | Ref. | Summary | Section |
|---------|------|---------|---------|
| HUA | [115] | RPCA standard version | IV-B |
| RAF1 | [130] | REPET standard version | IV-C |
| RAF2 | [134] | REPET with time-varying period | |
| RAF3 | [135] | REPET with similarity matrix | |
| KAM1-2 | [142] | KAM with different configurations | |
| CHA | [162] | RPCA with vocal activation information | V-A |
| JEO1-2 | [163] | $l_1$-RPCA with vocal activation information | |
| DUR | [201] | Source-filter NMF | V-C |
| OZE | [212] | Structured NMF with learned dictionaries | VI-B |
| KON | [291] | RNN | VI-C |
| GRA2-3 | [308] | DNN ensemble | |
| STO1-2 | [363] | FNN on *common fate* TF representation | |
| UHL1 | [293] | FNN with context | |
| NUG1-4 | [343] | FNN with multichannel information | VII |
| UHL2-3 | [294] | LSTM with multichannel information | |
| IBM | | ideal binary mask | |

The objective scores for these methods were obtained using BSS Eval and are given in Figure 12. For more details about the results and for listening to the estimates, we refer the reader to the dedicated interactive website[3].

As we first notice in Figure 12, the HUA method, corresponding to the standard RPCA as discussed in Section IV-B, showed rather disappointing performance in this evaluation. After inspection of the results, it appears that processing full-length tracks is the issue there: at such scales, vocals also exhibit redundancy, which is captured by the low-rank model associated with the accompaniment. On the other hand, the RAF1-3 and KAM1-3 methods that exploit redundancy through repetitions, as presented in Section IV-C, behave much better for full-length tracks: even if somewhat redundant, vocals are rarely as repetitive as the accompaniment. When those methods are evaluated on datasets with very short excerpts (e.g., MIR-1K), such severe practical drawbacks are not apparent.

Likewise, the DUR method that jointly models the vocals as harmonic and the accompaniment as redundant, as discussed in Section V-C, does show rather disappointing performance, considering that it was long the state-of-the-art in earlier SiSECs [270]. After inspection, we may propose two reasons for this performance drop. First, using full-length excerpts also clearly revealed a shortcoming of the approach: it poorly handles silences in the lead, which were rare in the short-length excerpts tested so far. Second, using a much larger evaluation set revealed that vocals are not necessarily well modeled by a harmonic source-filter model; breathy or saturated voices appear to greatly challenge such a model.

---

[3] http://www.sisec17.audiolabs-erlangen.de

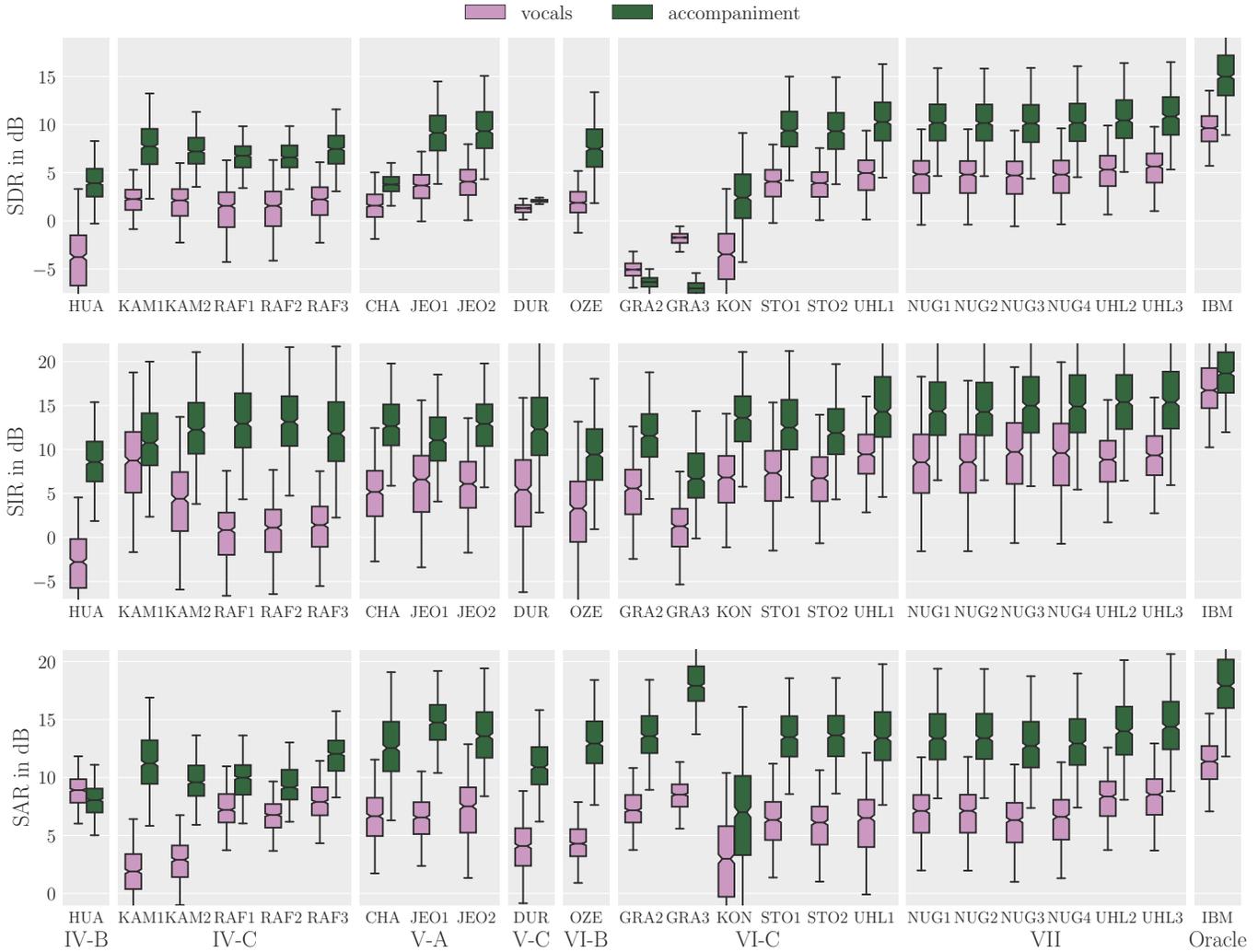

Fig. 12: BSS Eval scores for the vocals and accompaniment estimates for SiSEC 2016 on the DSD100 dataset. Results are shown for the *test* set only. Scores are grouped as in Table II according to the section they are described in the text, indicated below each group.

While processing full-length tracks comes as a challenge, it can also be an opportunity. It is indeed worth noticing that whenever RPCA is helped through vocal activity detection, its performance is significantly boosted, as highlighted by the relatively good results obtained by CHAN and JEO.

As discussed in Section VI, the availability of learning data made it possible to build data-driven approaches, like the NMF-based OZE method which is available through the Flexible Audio Source Separation Toolbox (FASST) [211], [212]. Although it was long state-of-the-art, it has been strongly outperformed recently by other data-driven approaches, namely DNNs. One first reason clearly appears as the superior expressive power of DNNs over NMF, but one second reason could very simply be that OZE should be trained anew with the same large amount of data.

As mentioned above, a striking fact we see in Figure 12 is that the overall performance of data-driven DNN methods is the highest. This shows that exploiting learning data does help separation greatly compared to only relying on *a priori* assumptions such as the harmonicity or redundancy. Additionally, dynamic models such as CNN or LSTM appear more adapted to music than FNN. These good performances in audio source separation go in line with the recent success of DNNs in fields as varied as computer vision, speech recognition, and natural language processing [285].

However, the picture may be seen to be more subtle than simply black-box DNN systems beating all other approaches. For instance, exploiting multichannel probabilistic models, as discussed in Section VII, leads to the NUG and UHL2-3 methods, that significantly outperform the DNN methods ignoring stereo information. In the same vein, we expect other specific assumptions and musicological ideas to be exploited for further improving the quality of the separation.

One particular feature of this evaluation is that it also shows obvious weaknesses in the objective metrics. For instance, the GRA method behaves significantly worse than any other methods. However, when listening to the separated signals, this does not seem deserved. All in all, designing new and convenient metrics that better match perception and that are specifically built for music on large datasets clearly appears as a desirable milestone.

In any case, the performance achieved by a totally informed filtering method such as IBM is significantly higher than that of any submitted method in this evaluation. This means that lead and accompaniment separation has room for much improvement, and that the topic is bound to witness many breakthroughs still. This is even more true considering that IBM is not the best upper bound for separation performance: other filtering methods such as *ideal ratio mask* [20] or multi-channel Wiener filter [344] may be considered as references.

Regardless of the above, we would also like to highlight that good algorithms and models can suffer from slight errors in their low-level audio processing routines. Such routines may include the STFT representation, the overlap-add procedure, energy normalization, and so on. Considerable improvements may also be obtained by using simple tricks and, depending on the method, large impacts can occur in the results by only changing low-level parameters. These include the overlap ratio for the STFT, specific ways to regularize matrix inverses in multichannel models, etc. Further tricks such as the exponentiation of the TF mask by some positive value can often boost performance significantly more than using more sophisticated models. However, such tricks are often lost when publishing research focused on the higher-level algorithms. We believe this is an important reason why sharing source code is highly desirable in this particular application. Some online repositories containing implementations of lead and accompaniment separation methods should be mentioned, such as **nussl**[4] and **untwist** [364]. In the companion webpage of this paper[5], we list many different online resources such as datasets, implementations, and tools that we hope will be useful to the practitioner and provide some useful pointers to the interested reader.

### D. Discussion

Finally, we summarize the core advantages and disadvantages for each one of the five groups of methods we identified.

Methods based on the harmonicity assumption for the lead are focused on sinusoidal modeling. They enjoy a very strong interpretability and allow for the direct incorporation of any prior knowledge concerning pitch. Their fundamental weakness lies in the fact that many singing voice signals are not harmonic, e.g., when breathy or distorted.

Modeling the accompaniment as redundant allows to exploit long-term dependencies in music signals and may benefit from high-level information like tempo or score. Their most important drawback is to fall short in terms of voice models: the lead signal itself is often redundant to some extent and thus partly incorporated in the estimated accompaniment.

Systems jointly modeling the lead as harmonic and the accompaniment as redundant benefit from both assumptions. They were long state-of-the-art and enjoy a good interpretability, which makes them good candidates for interactive separa-tion methods. However, their core shortcoming is to be highly sensitive to violations of their assumptions, which proves to often be the case in practice. Such situations usually require fine-tuning and hence prevents their use as black-box systems for a broad audience.

Data-driven methods involve machine learning to directly learn a mapping between the mixture and the constitutive sources. Such a strategy recently introduced a breakthrough compared to everything that was done before. Their most important disadvantages are the lack of interpretability, which makes it challenging to design good user interactions, as well as their strong dependency on the size of the training data.

Finally, multichannel methods leverage stereophonic information to strongly improve performance. Interestingly, this can usually be combined with better spectrogram models such as DNNs to further improve quality. The price to pay for this boost in performance is an additional computational cost, that may be prohibitive for recordings of more than two channels.

## IX. CONCLUSION

In this paper, we thoroughly discussed the problem of separating lead and accompaniment signals in music recordings. We gave a comprehensive overview of the research undertaken in the last 50 years on this topic, classifying the different approaches according to their main features and assumptions. In doing so, we showed how one very large body of research can be described as being model-based. In this context, it was evident from the literature that the two most important assumptions behind these models are that the lead instrument is harmonic, while the accompaniment is redundant. As we demonstrated, a very large number of methods on model-based lead-accompaniment separation can be seen as using one or both of these assumptions.

However, music encompasses a variety of signals of an extraordinary diversity, and no rigid assumption holds well for all signals. For this reason, while there are often some music pieces where each method performs well, there will also be some where it fails. As a result, data-driven methods were proposed as an attempt to introduce more flexibility at the cost of requiring representative training data. In the context of this paper, we proposed the largest freely available dataset for music separation, comprising close to 10 hours of data, which is 240 times greater than the first public dataset released 10 years ago.

At present, we see a huge focus on research utilizing recent machine learning breakthroughs for the design of singing voice separation methods. This came with an associated boost in performance, as measured by objective metrics. However, we have also discussed the strengths and shortcomings of existing evaluations and metrics. In this respect, it is important to note that the songs used for evaluation are but a minuscule fraction of all recorded music, and that separating music signals remains the processing of an artistic means of expression. As such it is impossible to escape the need for human perceptual evaluations, or at least adequate models for it.

After reviewing the large existing body of literature, we may conclude here by saying that lead and accompaniment separation in music is a problem at the crossroads of many different paradigms and methods. Researchers from very different backgrounds such as physics, signal or computer engineering have tackled it, and it exists both as an area for strong theoretical research and as a real-world challenging engineering problem. Its strong connections with the arts and digital humanities have proved attractive to many researchers.

---

[4]https://github.com/interactiveaudiolab/nussl
[5]https://sigsep.github.io

Finally, as we showed, there is still much room for improvement in lead and accompaniment separation, and we believe that new and exciting research will bring new breakthroughs in this field. While DNN methods represent the latest big step forward and significantly outperform previous research, we believe that future improvements can come from any direction, including those discussed in this paper. Still, we expect future improvements to initially come from improved machine learning methodologies that can cope with reduced training sets, as well as improved modeling of the specific properties of musical signals, and the development of better signal representations.

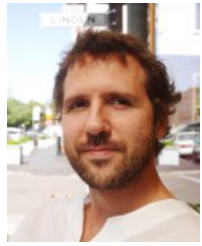

**Antoine Liutkus** received the State Engineering degree from Télécom ParisTech, France, in 2005, and the M.Sc. degree in acoustics, computer science and signal processing applied to music (ATIAM) from the Université Pierre et Marie Curie (Paris VI), Paris, in 2005. He worked as a research engineer on source separation at Audionamix from 2007 to 2010 and obtained his PhD in electrical engineering at Télécom ParisTech in 2012. He is currently researcher at Inria, France. His research interests include audio source separation and machine learning.

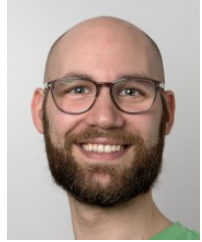

**Fabian-Robert Stöter** received the diploma degree in electrical engineering in 2012 from the Leibniz Universität Hannover and worked towards his Ph.D. degree in audio signal processing in the research group of B. Edler at the International Audio Laboratories Erlangen, Germany. He is currently researcher at Inria, France. His research interests include supervised and unsupervised methods for audio source separation and signal analysis of highly overlapped sources.

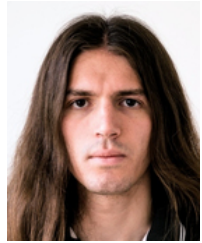

**Stylianos Ioannis Mimilakis** received a Master of Science degree in Sound & Music Computing from Pompeu Fabra University and a Bachelor of Engineering in Sound & Music Instruments Technologies from Higher Technological Education Institute of Ionian Islands. Currently he is pursuing his Ph.D. in signal processing for music source separation, under the MacSeNet project at Fraunhofer IDMT. His research interests include, inverse problems in audio signal processing and synthesis, singing voice separation and deep learning.

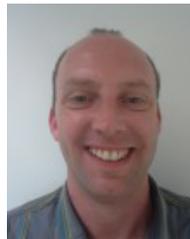

**Derry FitzGerald** (PhD, M.A. B.Eng.) is a Research Felow in the Cork School of Music at Cork Institute of Technology. He was a Stokes Lecturer in Sound Source Separation algorithms at the Audio Research Group in DIT from 2008-2013. Previous to this he worked as a post-doctoral researcher in the Dept. of Electronic Engineering at Cork Institute of Technology, having previously completed a Ph.D. and an M.A. at Dublin Institute of Technology. He has also worked as a Chemical Engineer in the pharmaceutical industry for some years. In the field of music and audio, he has also worked as a sound engineer and has written scores for theatre. He has utilised his sound source separation technologies to create the first ever officially released stereo mixes of several songs for the Beach Boys, including Good Vibrations and I get around. His research interests are in the areas of sound source separation and, tensor factorizations.

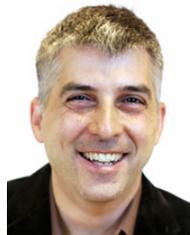

**Bryan Pardo,** head of the Northwestern University Interactive Audio Lab, is an associate professor in the Northwestern University Department of Electrical Engineering and Computer Science. Prof. Pardo received a M. Mus. in Jazz Studies in 2001 and a Ph.D. in Computer Science in 2005, both from the University of Michigan. He has authored over 80 peer-reviewed publications. He has developed speech analysis software for the Speech and Hearing department of the Ohio State University, statistical software for SPSS and worked as a machine learning researcher for General Dynamics. While finishing his doctorate, he taught in the Music Department of Madonna University. When he's not programming, writing or teaching, he performs throughout the United States on saxophone and clarinet at venues such as Albion College, the Chicago Cultural Center, the Detroit Concert of Colors, Bloomington Indiana's Lotus Festival and Tucson's Rialto Theatre.

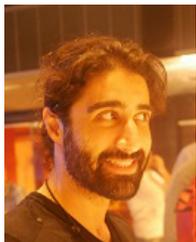

**Zafar Rafii** received a PhD in Electrical Engineering and Computer Science from Northwestern University in 2014, and an MS in Electrical Engineering from both Ecole Nationale Supérieure de l'Electronique et de ses Applications in France and Illinois Institute of Technology in the US in 2006. He is currently a senior research engineer at Gracenote in the US. He also worked as a research engineer at Audionamix in France. His research interests are centered on audio analysis, somewhere between signal processing, machine learning, and cognitive science, with a predilection for source separation and audio identification in music.